\documentclass[%
pre,
preprint,
tightenlines,
a4paper,12pt
]{revtex4}
\usepackage{dcolumn}
\usepackage{amssymb,amsmath,array}
\usepackage{graphicx}
\usepackage{subfigure}

\renewcommand{\thesubfigure}{(\alph{subfigure})}

\makeatletter
\renewcommand{\@thesubfigure}{\thesubfigure\space}

\newcommand{\bs}[1]{\boldsymbol{#1}}
\newcommand{\vc}[1]{\mathbf{#1}}
\newcommand{\uvc}[1]{\mathbf{\hat #1}}

\newcommand{\dd}{\mathrm{d}}
\newcommand{\lc}{\mathrm{N}}
\newcommand{\azo}{\mathrm{A}}

\newcommand{\iso}{\mathrm{iso}}
\newcommand{\scal}{\mathrm{scal}}
\newcommand{\av}{\mathrm{av}}
\newcommand{\LG}{\mathrm{LG}}
\newcommand{\exc}{\mathrm{ex}}

\newcommand{\sca}[2]{\bigl({#1}\cdot{#2}\bigr)}
\newcommand{\avr}[1]{\left\langle{#1}\right\rangle}

\newcommand{\degc}{$^\circ$C}

\newcommand{\Tr}{\mathop{\rm Tr}\nolimits}

\makeatother

\begin{document}
\DeclareGraphicsExtensions{.jpg,.pdf}
\title{Photo-induced ordering and anchoring properties of azo-dye films}

\author{Alexei~D.~Kiselev}
\altaffiliation[Present address: ]{%
 Institute of Physics of NASU,
 prospekt Nauki 46,
 03028 Ky\"{\i}v, Ukraine} 

\email[Email address: ]{kisel@mail.cn.ua}

 \author{Vladimir Chigrinov}
 \email[Email address: ]{eechigr@ust.hk}

\author{Dan Ding Huang}
 \email[Email address: ]{eehdx@ust.hk}
 \affiliation{%
 Hong Kong University of Science and Technology,
 Clear Water Bay, Kowloon, Hong Kong
 }

\date{\today}

\begin{abstract}
We study both theoretically and experimentally 
anchoring properties of photoaligning azo-dye films
in contact with a nematic liquid crystal depending on photo-induced
ordering of azo-dye molecules.
In the mean field approximation, we found that 
the bare surface anchoring energy
linearly depends on the azo-dye order parameter and
the azimuthal anchoring strength decays to zero in the limit of
vanishing photo-induced ordering.
From the absorption dichroism spectra measured in
the azo-dye films that are prepared from the azo-dye derivative with
polymerizable terminal groups (SDA-2) we obtain 
dependence of the dichroic ratio on the irradiation dose.
We also measure the polar and azimuthal anchoring
strengths in nematic liquid crystal (NLC) cells aligned by the azo-dye films
and derive the anchoring strengths as functions of the dichroic ratio
which is proportional to the photo-induced order parameter.
Though linear fitting of the experimental data for both anchoring
strengths gives reasonably well results, it, in contradiction with the
theory,  predicts vanishing of
the azimuthal anchoring strength at certain non-zero
value of the azo-dye order parameter.
By using a simple phenomenological model
we show that this discrepancy can be attributed to the difference
between the surface and bulk order parameters in the films.    
The measured polar anchoring energy is found to be an order of magnitude
higher than the azimuthal strength.
Our theory suggests that 
the quadrupole term of the spherical harmonics expansion
for the azo-dye~--~NLC intermolecular potential might be of importance for
the understanding of this difference.
\end{abstract}

 
\maketitle

\section{Introduction}
\label{sec:intro}

When a nematic liquid crystal (NLC) is brought into contact with an
anisotropic substrate, the energy of the NLC molecules in the
interfacial layer and thus the surface tension 
[the excess free energy per unit area]
will be orientationally
dependent.  The anisotropic part of the surface tension~---~the
so-called \textit{anchoring energy}~---~gives rise to the phenomenon
known as \textit{anchoring}, i.e., surface induced 
alignment of the nematic director along the vector of preferential orientation
referred to as the \textit{easy axis}.

Over the past few decades anchoring properties of NLCs have been the
subject of intense studies for both technological and more fundamental
reasons. There are a number of surface ordering and
anchoring transitions that were observed experimentally and 
were studied using different theoretical approaches
[see, \textit{e.g.},  Refs.~\cite{Sluck:in:1986,Jerom:rpp:1991,Durand:1996} for reviews].

Technologically, producing substrates with anisotropic anchoring
properties is of vital importance in the fabrication of liquid crystal
electrooptic devices. The traditional technique
widely used to align liquid crystal display cells involves mechanical
rubbing of aligning layers. This method, however, 
has the well known difficulties related to physical damage,
impurities, dust contamination and generation of electrostatic charge~\cite{Chigr:1999}.

An alternative \textit{photoalignment} technique avoiding
the drawbacks of the mechanical surface treatment 
was suggested in Refs.~\cite{Gibbon:nat:1991,Chig:jjap:1992,Dyad:jetpl:1992}. 
It uses linearly
polarized ultraviolet (UV) light to induce anisotropy of the angular
distribution of molecules in a photosensitive film~\cite{Kelly:jpd:2000}.

The photoalignment has been extensively studied in a number of
different polymer systems such as dye doped polymer layers~\cite{Gibbon:nat:1991,Furum:1999},
cinnamate polymer
derivatives~\cite{Chig:jjap:1992,Dyad:jetpl:1992,Gal:1996,Barn:2000,Yar:2002}
and side chain
azopolymers~\cite{Petry:1993,Holme:1996,Blin:1998,Iked:2000,Kis:epj:2001,Yar:2001,Kis:pre:2003}.
Light induced ordering  in the
photosensitive materials, though not being understood very well,
can occur by a variety of photochemically induced
processes.  These typically may involve such transformations as photoisomerization, crosslinking,
photodimerization and photodecomposition 
(a recent review can be found in Ref.~\cite{Chigr:rewiev:2003}).

In this paper we examine anchoring properties of the films containing photochemically stable azo dye
structures that were recently studied as  new photoaligning materials for NLC 
cells~\cite{Chig:lc:2002,Chig:pre:2003}. 
Dependence of the surface anchoring strengths on the photoinduced
anisotropy will be of our primary interest.

More specifically, we are aimed to study the effects of the photoinduced ordering
in azo-dye films on the polar and azimuthal anchoring energies.
The key point is that the photoalignment technique provides a means for controlling 
the photoinduced ordering that affects anchoring properties of
photoaligning layers by  changing ordering of azo-dye
molecules at the surface  and, thus, the surface anchoring strengths.

Recently,
the anchoring properties of aligning photopolymer layers 
in relation to the photoinduced ordering were studied experimentally  
in Ref.~\cite{Thie:pre:2003}. The relationship between the rubbing
strength and the azimuthal anchoring energy was discussed in Ref.~\cite{Oka:pre:2004}

The photopolymer-NLC interface was also described theoretically
in Refs.~\cite{Kis:ufz:2001,Alexe:pre:2001}
using a modified version of the variational mean field approach
which is also known as the Maier-Saupe theory.
 By contrast, the azo-dye films have not yet received a proper attention 
and we intend to fill in the gap.

The paper is organized as follows.
In  Sec.~\ref{sec:theory}
we apply the mean field theoretical 
approach~\cite{Sen:1987,Sluck:jcp1:1992,Sluck:jcp2:1992} 
to express the surface anchoring energy in terms of the tensorial order parameters
which characterize angular distribution of the azo-dye and NLC molecules at 
the interfacial boundary surface. The general result is then used to
derive the expressions for the azimuthal and polar anchoring strengths
that, in addition to the order parameters, depend on the harmonics of
the intermolecular potentials.

Experimental details are given in Sec.~\ref{sec:experim}.
The polymerizable azo-dye monomer SDA-2 was used to prepare 
the photoaligning layers. Absorption dichroism spectra were measured
in the films irradiated with linearly polarized UV light at various
irradiation doses. Anchoring energy measurements were performed in
NLC cells where NLC is sandwiched between the glass plates coated with
the azo-dye film. 

In Sec.~\ref{sec:results} we present the experimental results
and apply the theory of Sec.~\ref{sec:theory} to interpret the data.
Discussion and concluding remarks are given in
Sec.~\ref{sec:discussion}.
Details on some technical results are relegated to 
appendices~\ref{sec:order-tensor}-\ref{sec:spher-harm-expans}.
 
\section{Theory}
\label{sec:theory}

In this section we begin with introducing  general notations 
and apply the mean-field approach to express the Landau-de~Gennes
surface free energy in terms of both azo-dye and NLC order parameters.
Expressions for the azimuthal and polar anchoring strengths,
$W_{\phi}$ and $W_{\theta}$, are then derived from the orientationally
dependent part of the surface energy in Sec.~\ref{subsec:anch-energy}.
In the concluding part of this section we consider effects of spatial
variations of the azo-dye order parameter using a simple model  
formulated in Sec.~\ref{subsec:spat-vari-order}.

\subsection{Order parameter and dichroic ratio}
\label{sec:order-param-dich}

Assuming that the unit vector, 
$\uvc{u}=(\sin\theta\cos\phi,\sin\theta\sin\phi,\cos\theta)$,
directed along the long molecular axis defines
orientation of a molecule in both azo-dye film and  NLC cell,
quadrupolar orientational ordering of the molecules 
can be characterized using
the  traceless symmetric second-rank tensor~\cite{Gennes:bk:1993}
\begin{equation}
  \label{eq:Q-def}
  \vc{Q}(\uvc{u})=(3\,\uvc{u}\otimes\uvc{u}-\vc{I})/2,
\end{equation}
where $\vc{I}$ is the identity matrix.
The dyadic~\eqref{eq:Q-def} averaged 
over orientation of molecules with 
the one-particle distribution function 
$\rho_{\alpha}(\vc{r},\uvc{u})$,
describing the orientation-density
profile of azo-dye ($\alpha=\azo$) and NLC ($\alpha=\lc$) molecules,
is proportional to the \emph{order parameter tensor} $\vc{S}_{\alpha}(\vc{r})$ 
\begin{align}
  \label{eq:avr-rho}
  \int\rho_{\alpha}(\vc{r},\uvc{u})\vc{Q}(\uvc{u})\dd\uvc{u}=
\rho_{\alpha}(\vc{r})\vc{S}_{\alpha}(\vc{r}),
\end{align}
where $\dd\uvc{n}\equiv\sin\theta\dd\theta\dd\phi$,
$\rho_{\alpha}(\vc{r},\uvc{u})=\rho_{\alpha}(\vc{r}) f_{\alpha}(\vc{r},\uvc{u})$,
$\rho_{\alpha}(\vc{r})=\int\rho_{\alpha}(\vc{r},\uvc{u})\dd\uvc{u}$
is the density profile and
$f_{\alpha}(\vc{r},\uvc{u})$ is the normalized angular distribution.
The general expression
for the order parameter is given in appendix~\ref{sec:order-tensor}  
[see Eq.~\eqref{eq:avr-Q}] along with technical details on 
the technique of irreducible tensors. 

Now we dwell briefly on the relation between the
order parameter $\vc{S}_{\azo}$ characterizing orientational
distribution of azo-dye molecules $f_{\azo}(\uvc{u})$ 
and the absorption dichroic ratio
\begin{equation}
  \label{eq:dichroic}
  R=\frac{D_{\parallel}-D_{\perp}}{D_{\parallel}+2 D_{\perp}},
\end{equation}
where $D_{\parallel}$ [$D_{\perp}$] is the absorption coefficient measured 
for a testing beam linearly polarized parallel
[perpendicular] to the polarization vector of the activating UV
light which is directed along the $x$ axis, $\vc{E}_{\exc}= E_{\exc}\,\uvc{x}$.
We shall also assume that
the testing and the pumping waves are both propagating along the $z$
axis which is normal to the film substrate.  

When the absorption tensor of an azo-dye molecule is uniaxially
anisotropic
with $\sigma_{ij}(\uvc{u})=\sigma_{\perp}\delta_{ij}+
(\sigma_{\parallel}-\sigma_{\perp}) u_i u_j$,
its orientational average takes the following matrix form
\begin{align}
&
  \label{eq:sigma_avr}
\avr{\bs{\sigma}}=\bigl(\sigma_{\av}\, \vc{I}+2\,
  \Delta\sigma\,\vc{S}_{\azo}\bigr)/3,
\\
&
\label{eq:sigma_param}
\sigma_{\av}=\sigma_{\parallel}+2\sigma_{\perp},  
\quad
\Delta\sigma=\sigma_{\parallel}-\sigma_{\perp},
\end{align}
where the angular brackets $\avr{\dots}$ denote orientational averaging.

In the low concentration approximation, the optical densities 
$D_{\parallel}$ and $D_{\perp}$
are
proportional to the corresponding components of the tensor~\eqref{eq:sigma_avr}
\begin{align}
&
  \label{eq:D_par}
  D_{\parallel}\propto \rho_{\azo}
\bigl(\sigma_{\av}+2\,
  \Delta\sigma\,S_{xx}^{(\azo)}\bigr)/3,
\\
&
  \label{eq:D_perp}
  D_{\perp}\propto \rho_{\azo}
\bigl(\sigma_{\av}+2\,
  \Delta\sigma\,S_{yy}^{(\azo)}\bigr)/3,
\end{align}
so that the average
absorption coefficient $D_{\av}$ is given by
\begin{equation}
  \label{eq:D_avr}
  D_{\av}=D_{\parallel}+2D_{\perp}\propto \rho_{\azo}\bigl(
\sigma_{\av}+
2/3\,
  \Delta\sigma\,\left[S_{yy}^{(\azo)}-S_{zz}^{(\azo)}\right]
\bigr).
\end{equation}
When the absorption coefficient $D_{\av}$ does not depend on 
irradiation dose (and, thus, on the order parameter),
from the expression~\eqref{eq:D_avr} 
we may conclude that anisotropy of the azo-dye film is uniaxial and 
$S_{yy}^{(\azo)}=S_{zz}^{(\azo)}=-S_{xx}^{(\azo)}/2 \equiv -S_{\azo}/2$.  
In this case we have
\begin{equation}
  \label{eq:dichroic2}
\vc{S}_{\azo}=S_{\azo} (3\,\uvc{x}\otimes\uvc{x}-\vc{I})/2,\quad
  R=\frac{\Delta\sigma}{\sigma_{\av}}\,S_{\azo}.
\end{equation}

As is seen from Eq.~\eqref{eq:dichroic2}, the dichroic ratio equals
the order parameter only in the limiting case 
where absorption  of waves propagating along
the long molecular axis is negligibly small and $\sigma_{\perp}\to 0$.

\subsection{ Anisotropic part of surface energy in the mean-field approximation}
\label{subsec:mean-field}

In the previous section it was shown that 
the light induced ordering of azo-dye molecules
can be described
by the order parameter~\eqref{eq:dichroic2} which is expected to affect
the surface free energy at the nematic-substrate interface.
So, in this section, the order parameter dependent part of the surface
energy will be of our primary concern.

In the case of a flat structureless substrate, 
the expression for the surface energy
was originally obtained by Sen and Sullivan in Ref.~\cite{Sen:1987}. 
Subsequently, similar results have been derived
by using the mean-field approximation~\cite{Sluck:jcp1:1992} 
and the density functional theory~\cite{Hess:jcp:1993,Sluck:jpf:1993,Sluck:pre:1997}.

Similarly to Ref.~\cite{Sluck:jcp1:1992} , we adopt the mean-field
approach and use the Fowler approximation for the one-particle
distribution functions
\begin{align}
  \label{eq:Fowler}
  \rho_{\lc}(\vc{r},\uvc{u})= H(z) \rho_{\lc}(z,\uvc{u}),
\quad
  \rho_{\azo}(\vc{r},\uvc{u})= H(-z) \rho_{\azo}(z,\uvc{u}),
\end{align}
where $H(z)$ is the Heaviside step function which equals unity when
$z$ is positive and vanishes otherwise.

Applying the mean-field theory~\cite{Sluck:jcp1:1992} gives the 
Landau-de Gennes surface free energy as an
excess Helmholtz free energy per unit area that depends on
two pair intermolecular potentials:
(a)~the potential of interaction between NLC molecules, 
$U_{\lc-\lc}(\vc{r}_{12},\uvc{u}_1,\uvc{u}_2)$; and 
(b)~the potential of interaction between NLC and azo-dye molecules, 
$U_{\azo-\lc}(\vc{r}_{12},\uvc{u}_1,\uvc{u}_2)$,
where $\vc{r}_{12}=\vc{r}_1-\vc{r}_2$ is the vector of intermolecular
separation and $\uvc{u}_i$ is the orientation coordinates of the
interacting molecules. 
The resulting expression is given by
\begin{align}
&
  \label{eq:delta-F}
  \Delta F/A=
\int_{-\infty}^{0} \dd z_1\int_{0}^{\infty}\dd z_2\int\dd \uvc{u}_1\dd
\uvc{u}_2
\notag
\\
&
\times
\Bigl[\rho_{\azo}(z_1,\uvc{u}_1) V_{\azo}(z_{12},\uvc{u}_1,\uvc{u}_2)
\rho_{\lc}(z_2,\uvc{u}_2)
\notag
\\
&
-\frac{1}{2}\,
\rho_{\lc}(z_1,\uvc{u}_1) V_{\lc}(z_{12},\uvc{u}_1,\uvc{u}_2)
\rho_{\lc}(z_2,\uvc{u}_2)
\Bigr],
\\
&
  \label{eq:V-alph}
  V_{\alpha}(z_{12},\uvc{u}_1,\uvc{u}_2) =
\int_{A}
U_{\alpha-\lc}(\vc{r}_{12},\uvc{u}_1,\uvc{u}_2)
\dd x_{12}\dd y_{12},
\end{align}
where $A$ is the area of the substrate and $V_{\alpha}$ is the
potential averaged over in-plane coordinates.

It should be noted that the potentials $U_{\lc-\lc}$ and $U_{\azo-\lc}$ 
actually represent the perturbative part of
interaction that can be treated in the mean-field approximation.
They can be written in the form of expansion over spherical harmonics
given in Eq.~\eqref{eq:poten-gen} of
appendix~\ref{sec:spher-harm-expans}.
For our purposes, however, it is more convenient to use the tensorial
representation for the averaged potentials $V_{\alpha}$ that was
introduced in Ref.~\cite{Ron:pra:1980}.
In appendix~\ref{sec:spher-harm-expans}, the coefficients that enter
this representation [see Eq.~\eqref{eq:poten-in-Q}] are related to the
coefficients, $v_{j_1 j_2 j}(z)$ with $j_i<4$, 
in the spherical harmonics expansion~\eqref{eq:poten-in-Q}.
This relation is given by Eqs.~\eqref{eq:alph}--\eqref{eq:beta-3}.

Substituting the representation~\eqref{eq:poten-in-Q} into
Eq.~\eqref{eq:delta-F} and assuming homogeneity of the contacting
phases,
we obtain the Landau-de Gennes expression for the surface free energy in
the final form: 
\begin{align}
&
  \label{eq:f_s}
  f_S(\vc{S}_{\lc},\vc{S}_{\azo})=
f_{\lc}(\vc{S}_{\lc})+
+f_{\azo}(\vc{S}_{\lc},\vc{S}_{\azo}),
\end{align}
\begin{align}
  \label{eq:f_lc}
  f_{\lc}(\vc{S}_{\lc})&=
c_{0}\, \uvc{z}\cdot \vc{S}_{\lc}\cdot \uvc{z}+
c_{\lc}^{(1)} \Tr( \vc{S}_{\lc}^2)
\notag\\
&
+
c_{\lc}^{(2)}\,\uvc{z}\cdot \vc{S}_{\lc}^2\cdot \uvc{z}+
c_{\lc}^{(3)} \left[\uvc{z}\cdot \vc{S}_{\lc}\cdot \uvc{z}\right]^2,
\end{align}
\begin{align}
&
\label{eq:f_azo}
f_{\azo}(\vc{S}_{\lc},\vc{S}_{\azo})=
c_{\azo}^{(1)} 
\Tr( \vc{S}_{\lc}\,\vc{S}_{\azo})
\notag\\
&
+
c_{\azo}^{(2)}\, 
\uvc{z}\cdot \vc{S}_{\lc}\,\vc{S}_{\azo}\cdot \uvc{z}+
c_{\azo}^{(3)}
\left[\uvc{z}\cdot \vc{S}_{\lc}\cdot \uvc{z}\right]\cdot
\left[\uvc{z}\cdot \vc{S}_{\azo}\cdot \uvc{z}\right],
\end{align}
where the coefficients are given by
\begin{align}
  \label{eq:coef-c-0}
  c_0=b_{\azo}^{(0)}-b_{\lc}^{(0)},
\\
\label{eq:coef-c-i}
c_{\azo}^{(i)}=b_{\azo}^{(i)},\quad
c_{\lc}^{(i)}=-b_{\lc}^{(i)}/2,
\\
\label{eq:coef-b-i}
b_{\alpha}^{(i)}=
\rho_{\alpha}\rho_{\lc}\int_0^{\infty} z\beta_{\alpha}^{(i)}(z)\,\dd z,
\end{align}
$\beta_{\alpha}^{(i)}(z)$ denote the coefficients in the
representation~\eqref{eq:poten-in-Q} for the potential~\eqref{eq:V-alph}.

Eqs.~\eqref{eq:f_s}--\eqref{eq:f_azo} can be viewed as a generalization of
the expression by Sen and Sullivan~\cite{Sen:1987} supplemented with
the term $f_{\azo}(\vc{S}_{\lc},\vc{S}_{\azo})$ resulting from the
interaction between NLC and azo-dye molecules.  
Note that this result can also be derived by constructing
invariants from the order parameter tensors $\vc{S}_{\alpha}$
and the normal to the substrate $\uvc{z}$. 
In this case, the surface of the azo-dye aligning film is
treated phenomenologically as 
a bounding surface which, in addition to the normal $\uvc{z}$, is characterized
by the order parameter $\vc{S}_{\azo}$.

\subsection{Bare anchoring energy}
\label{subsec:anch-energy}

Separating out the director dependent part of the surface free energy
requires the order parameters of azo-dye and NLC molecules 
be substituted into Eqs.~\eqref{eq:f_s}--\eqref{eq:f_azo}.
Since the order parameter at the surface 
may differ from its value in the bulk, we generalize the expression
for the azo-dye order parameter~\eqref{eq:dichroic2} as follows  
\begin{align}
  \label{eq:S-azo-surf}
  2 \vc{S}_{\azo}\vert_{z=0} =
S_{\azo} (3 \uvc{x}\otimes\uvc{x}-\vc{I})+ 
P_{\azo} (\uvc{z}\otimes\uvc{z}-\uvc{y}\otimes\uvc{y}).
\end{align}
Similarly, for NLC order parameter tensor at $z=0$, 
from Eq.~\eqref{eq:avr-Q} we have
\begin{align}
&
\label{eq:S_N}
 2 \vc{S}_{\lc}\vert_{z=0} =
S (3 \uvc{n}\otimes\uvc{n}-\vc{I})+ 
P (\uvc{m}\otimes\uvc{m}-\uvc{l}\otimes\uvc{l})
\notag
\\
&
=(3S+P)\,\uvc{n}\otimes\uvc{n}+2P\, \uvc{m}\otimes\uvc{m} -(S+P)\,\vc{I}, 
\end{align}
where $\uvc{n}=(\sin\Theta\cos\Phi,\sin\Theta\sin\Phi,\cos\Theta)$
is the NLC director,
$\uvc{n}\perp\uvc{m}=\cos\gamma\,\vc{e}_x(\uvc{n})-\sin\gamma\,\vc{e}_y(\uvc{n})$,
$\vc{e}_x(\uvc{n})=(\cos\Theta\cos\Phi,\cos\Theta\sin\Phi,-\sin\Theta)$,
$\vc{e}_y(\uvc{n})=(-\sin\Phi,\cos\Phi,0)$
and $\uvc{l}=\uvc{n}\times\uvc{m}$.

Eqs.~\eqref{eq:S-azo-surf} and~\eqref{eq:S_N}  suggest that
the order parameters of azo-dye and NLC molecules
though being both uniaxial in the bulk can be biaxial at the surface.
In addition, the scalar order parameters
$S_{\azo}$ and $S$ at the surface may also deviate from their values in the bulk. 

The surface free energy can now be expressed as a sum of two contributions
\begin{align}
  \label{eq:f_s-sep}
  f_S(\vc{S},\vc{S}_{\azo})=
W(\uvc{n},\uvc{m})+f_{\scal},
\end{align}
where $W(\uvc{n},\uvc{m})$ 
is the orientationally dependent part of the surface energy.
This part can be calculated by substituting the order
parameters~\eqref{eq:S-azo-surf} and~\eqref{eq:S_N}
into the surface energy~\eqref{eq:f_s} to yield
the expression for the bare anchoring energy
\begin{align}
&
  \label{eq:f_anch}
 W(\uvc{n},\uvc{m})=N_{z}\sca{\uvc{n}}{\uvc{z}}^2+
M_z\sca{\uvc{m}}{\uvc{z}}^2
\notag
\\
&
+
N_x \sca{\uvc{n}}{\uvc{x}}^2
+M_x \sca{\uvc{m}}{\uvc{x}}^2
\notag
\\
&
+
c_{\lc}^{(3)}/4 \Bigl[
(3S+P)\,\sca{\uvc{n}}{\uvc{z}}^2
+2P\,\sca{\uvc{m}}{\uvc{z}}^2
\Bigr]^2
\end{align}
with the coefficients defined by the relations:
\begin{align}
&
  \label{eq:N-z}
  4 N_z=(3S+P)
\bigl[q+c_{\lc}^{(2)}(S-P)
\bigr],
\\
&
  \label{eq:N-x}
  4 N_x=c_{\azo}^{(1)} (3S_{\azo}+P_{\azo})(3S+P),
\\
&
\label{eq:M-zx}
  2 M_z=P
\bigl[
q-2c_{\lc}^{(2)} S
\bigr],\quad
  2 M_x=c_{\azo}^{(1)} (3S_{\azo}+P_{\azo}) P,
\\
&
\label{eq:q-aux}
q\equiv 2c_0-
\bigl(
c_{\azo}^{(2)}+c_{\azo}^{(3)}
\bigr)(S_{\azo}-P_{\azo})
+2 c_{\azo}^{(1)} P_{\azo} 
-2c_{\lc}^{(3)}(S+P).
\end{align}

The second term on the right hand side of Eq.~\eqref{eq:f_s-sep}
\begin{align}
&
  \label{eq:f_scal}
 4f_{\scal}=-\bigl[
2c_0-(S_{\azo}-P_{\azo})(c_{\azo}^{(2)}+c_{\azo}^{(3)})
+3c_{\azo}^{(1)}(S_{\azo}+P_{\azo}) 
\bigr]
\notag
\\
& 
\times (S+P)
+(c_{\lc}^{(2)}+c_{\lc}^{(3)})(S+P)^2
+2c_{\lc}^{(1)}(3S^2+P^2).
\end{align}
is a quadratic function of  NLC scalar order parameter $S$ and the biaxiality $P$.
From Eq.~\eqref{eq:f_scal} it, similarly to the anchoring energy~\eqref{eq:f_anch}, 
depends linearly on
the azo-dye parameters $S_{\azo}$ and $P_{\azo}$.

For the anchoring energy~\eqref{eq:f_anch}, we consider the simplest
case which occurs 
when the surface induced NLC biaxiality $P$ is negligibly small and 
the quadrupolar term $v_{224}$ in the expansion of the
intermolecular potential $V_{\lc}$ can be ignored.
Under these circumstances,
setting $P=c_{\lc}^{(3)}=0$ and $M_{z}=M_{x}=0$, 
we arrive at the simplified formula for the anchoring energy
\begin{equation}
  \label{eq:anch-simpl}
  W(\uvc{n})=N_{z}\sca{\uvc{n}}{\uvc{z}}^2+
N_x\sca{\uvc{n}}{\uvc{x}}^2
\end{equation}
which agrees with
the expression for the anchoring energy recently proposed in 
Refs.~\cite{Iwam:pre:2000,Zhao:2002,Chen:lc:2004}.

From Eq.~\eqref{eq:anch-simpl} it is clear that the easy axis is
directed along the $y$ axis, $\vc{e}_s=\uvc{y}$, only if the coefficients $N_z$ and $N_z$
are both positive. In this case the polar and the azimuthal anchoring strengths,
$W_{\theta}$ and $W_{\phi}$, are given by
\begin{equation}
  \label{eq:W-theta-phi}
  W_{\theta}=2 N_z,\quad W_{\phi}=2N_x.
\end{equation}

From Eq.~\eqref{eq:N-x} we immediately deduce a more explicit expression
for the azimuthal anchoring strength
\begin{align}
&
  \label{eq:wa-fit}
  W_{\phi}=w_{\phi} \bigl[
S_{yy}^{(\azo)}\vert_{z=0}-S_{xx}^{(\azo)}\vert_{z=0}
\bigr],
\\
&
\label{eq:coefa-fit}
2w_{\phi}=-3 c_{\azo}^{(1)} S\vert_{z=0},
\end{align}
where  notations indicate the plane $z=0$ as a surface 
separating the phases.

Similarly, Eq.~\eqref{eq:N-z}  gives the polar anchoring strength in
the explicit form
\begin{align}
&
  \label{eq:wp-fit}
  W_{\theta}=w_{\theta}^{(0)}+w_{\theta}^{(1)} S_{zz}^{(\azo)}\vert_{z=0}
-w_{\phi} \bigl[
S_{zz}^{(\azo)}\vert_{z=0}-S_{yy}^{(\azo)}\vert_{z=0}
\bigr],
\\
&
\label{eq:coefp-fit}
2w_{\theta}^{(1)}=3 \bigl(c_{\azo}^{(2)}+c_{\azo}^{(3)}\bigr)S\vert_{z=0}.
\end{align}

The formulas~\eqref{eq:wa-fit}-\eqref{eq:coefp-fit} will be
subsequently used in Sec.~\ref{sec:results} to interpret the
experimental data. At this stage,
it is worth noting that, for the order parameter~\eqref{eq:dichroic2}, 
the relations~\eqref{eq:coefa-fit} 
and~\eqref{eq:coefp-fit} provide the inequalities
\begin{equation}
  \label{eq:inequal}
  c_{\azo}^{(1)}<0,\quad
c_{\azo}^{(2)}+c_{\azo}^{(3)}>0
\end{equation}
as conditions for the anchoring
strengths to increase linearly  as the scalar order parameter
$S_{\azo}$ decreases.

\begin{figure*}[!tbh]
\centering
\resizebox{165mm}{!}{\includegraphics*{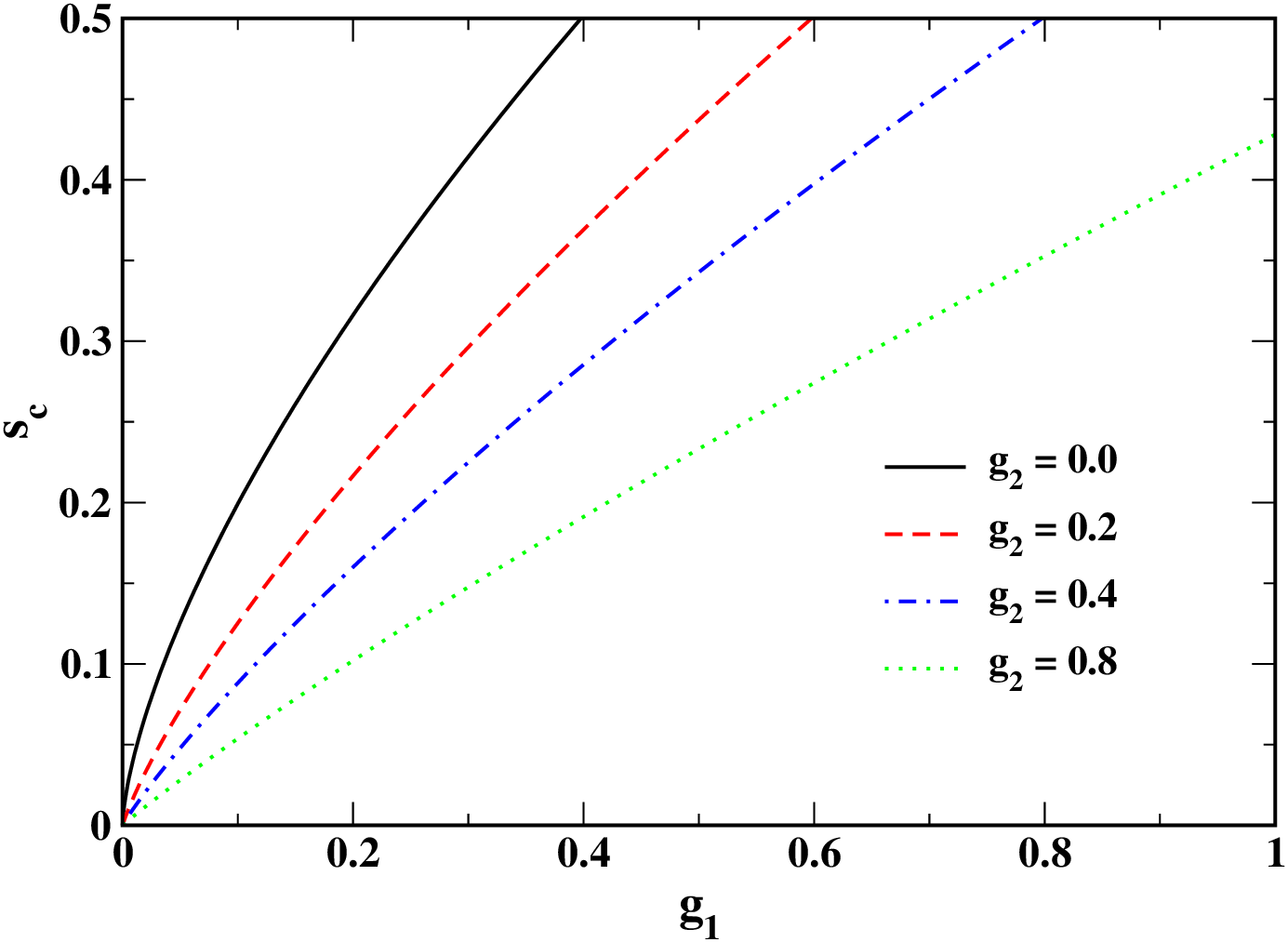}}
\caption{%
Critical order parameter $s_{c}$ as a function of the
coupling constant $g_1$ at various values of 
the coefficient $g_2$.
}
\label{fig:sc-g}
\end{figure*}

\begin{figure*}[!tbh]
\centering
\resizebox{165mm}{!}{\includegraphics*{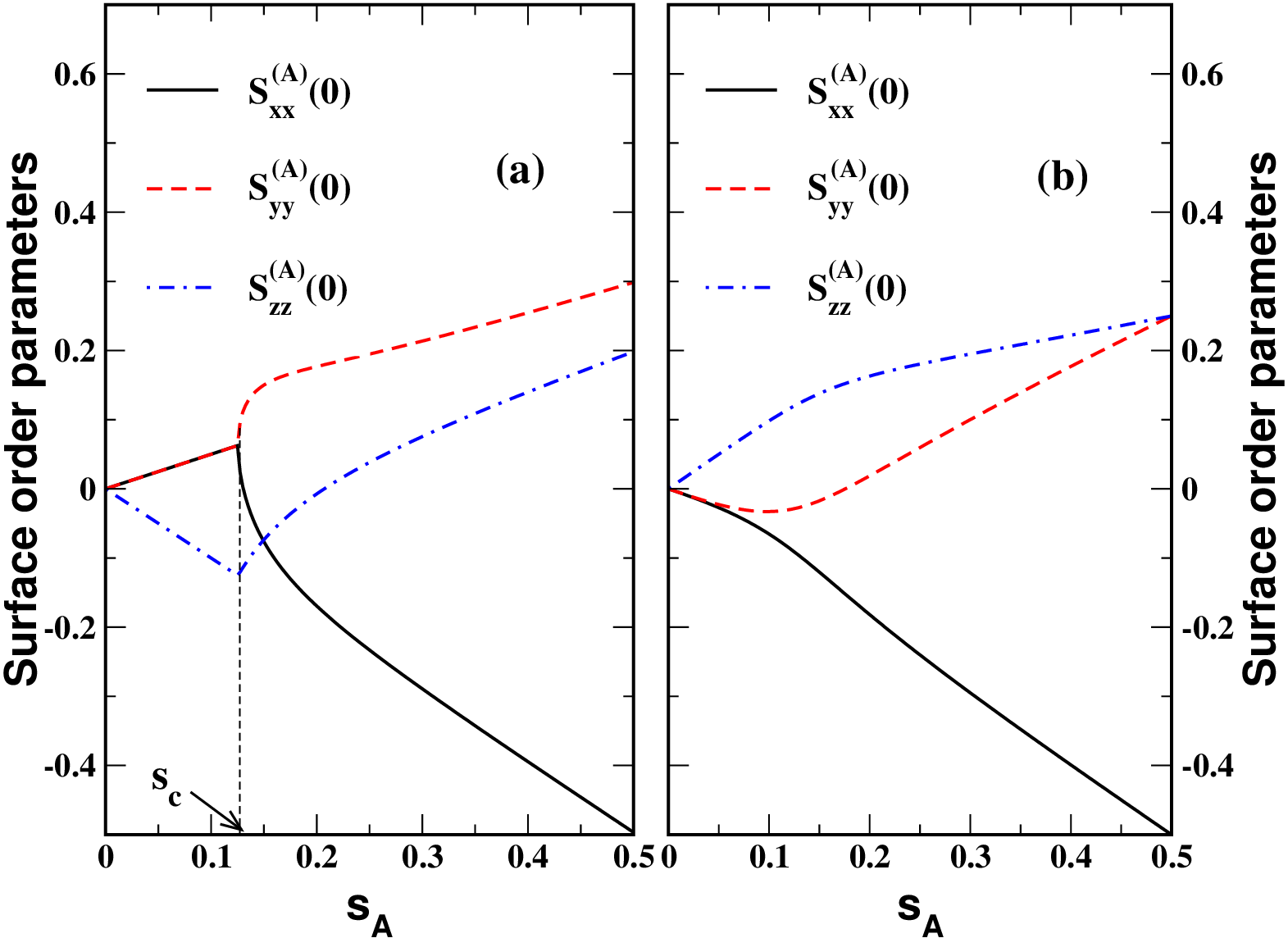}}
\caption{%
Components of the order parameter at the surface as functions
of the bulk order parameter $s_{\azo}$.
Two cases are shown:
(a)~$g_1=0.05$ and $g_2=0.0$;
(b)~$g_1=-0.2$ and $g_2=0.4$.
}
\label{fig:sii-sa}
\end{figure*}

\begin{figure*}[!tbh]
\centering
\resizebox{165mm}{!}{\includegraphics*{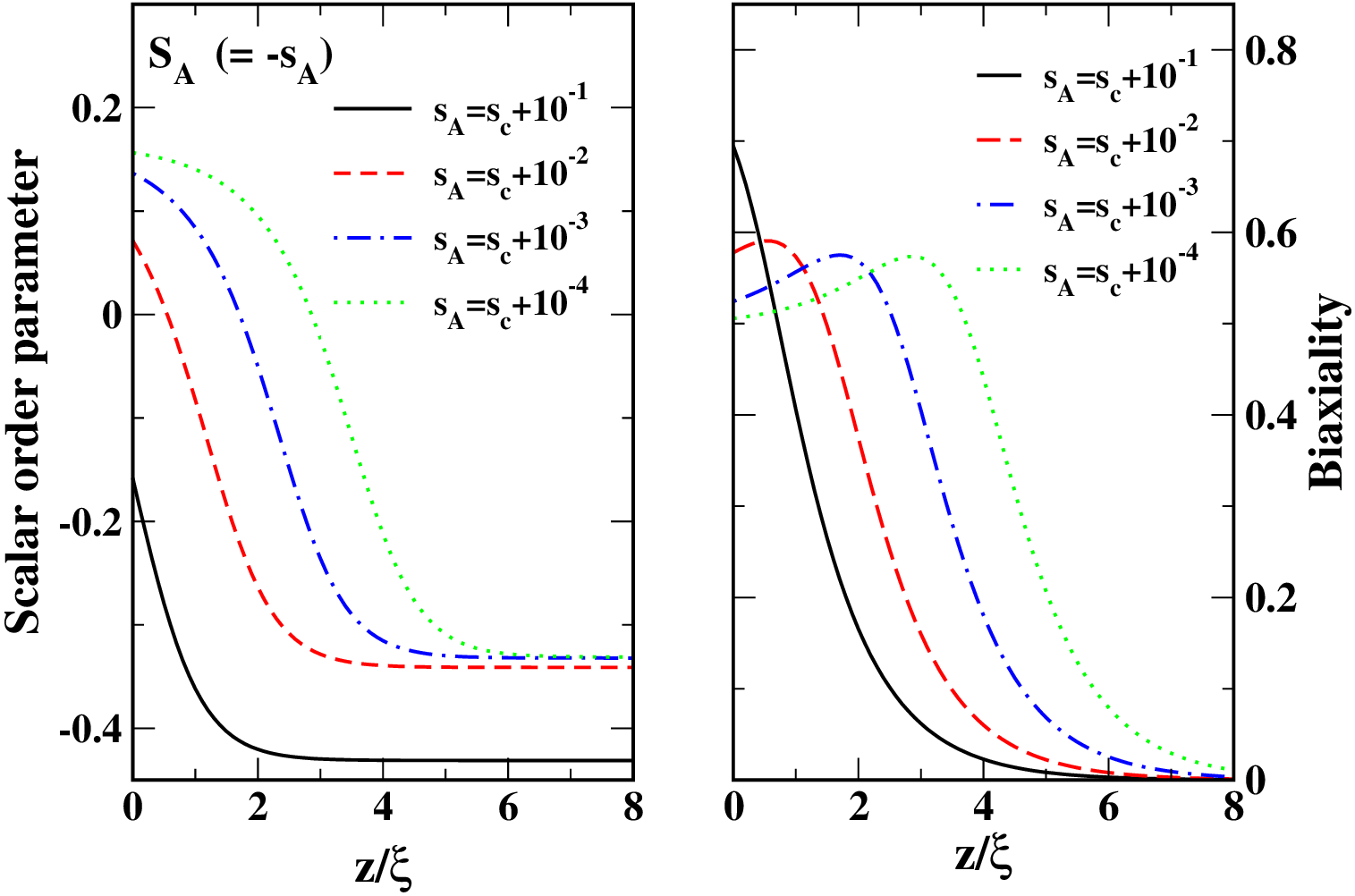}}
\caption{%
(a)~Scalar order parameter and (b)~biaxiality in relation to the
ratio $z/\xi$.
}
\label{fig:sp-z}
\end{figure*}

\subsection{Model of spatially varying order parameter}
\label{subsec:spat-vari-order}

As it was pointed out at the beginning of
the previous section, the surface order parameter tensor of
an azo-dye film~\eqref{eq:S-azo-surf} may differ from 
the bulk order parameter of the film~\eqref{eq:dichroic2}. 
The latter, according to the experimental results presented in the
subsequent section~\ref{sec:experim},
is uniaxially anisotropic with the in-plane anisotropy axis 
that is normal to the polarization vector of the activating UV light.
In addition,
the light induced scalar order parameter, which is proportional to the
dichroic ratio~\eqref{eq:dichroic}, turns out to be negative, $S_{\azo}^{(b)}<0$. 

From the other hand, assuming that the anisotropic part of the surface
energy can be taken in the general form by Sen and Sullivan~\cite{Sen:1987}, 
the boundary conditions may favor either homeotropic or
planar alignment of the azo-dye molecules, thus, counteracting the action of light.
So, it can be expected that
the effects caused by interplay between the light induced and the surface ordering 
are of importance in explaining the order parameter dependencies of
the polar and azimuthal anchoring energies.

In this section we discuss these effects on the basis of
a simple phenomenological model formulated by using the polar representation~\eqref{eq:polar-rep} for
the azo-dye order parameter. The latter can be conveniently rewritten
in the form 
\begin{equation}
  \label{eq:polar-azo}
  S_{\azo}=-s_{\azo}\cos\psi,\quad
P_{\azo}=-\sqrt{3}\,s_{\azo}\sin\psi,
\end{equation}
where the angle $\psi$ is shifted by $\pi$ so as to have the angle
$\psi$ vanishing in the bulk. 

In what follows we shall assume that, similarly to nematic liquid 
crystals~\cite{Lyuks:jetp:1978,Trebin:jpf:1989,Roso:jpa:1996,Kralj:jpa:2001}, 
the amplitude $s_{\azo}$ varies in space much slower than
the angle $\psi$. So, in our model, the amplitude will be
 fixed at its bulk value, $s_{\azo}=|S_{\azo}^{(b)}|$, and we consider
the limiting case of thick films in which the characteristic length of spatial variations of the angle
$\psi$ is much shorter than the film thickness. 
In this case the film can be regarded as a semi-infinite
sample filling the upper half space, $z\ge 0$.

Technically, our task will be to find the spatially varying angle $\psi$ as a function
of $z$ that minimizes the excess free energy per unit area,
$F_{\azo}$, taken in the following nematic-like form  
\begin{align}
&
  \label{eq:F-azo}
  F_{\azo}=
\int_0^{\infty}
\Bigl[
L s_{\azo}^2 \bigl(\partial_z \psi\bigr)^2+ B s_{\azo}^3 \bigl(1-\cos(3\psi)\bigr)
\Bigr]\dd z
\notag
\\
&
+ G_1\, s_{\azo}\cos(\psi_0+\pi/3)+G_2\, s_{\azo}^2 \cos^2(\psi_0+\pi/3),
\end{align}
where $\psi_0\equiv\psi\vert_{z=0}$ and 
$s_{\azo}\cos(\psi_0+\pi/3)=\uvc{z}\cdot
\vc{S}_{\azo}\cdot\uvc{z}\vert_{z=0}$.
The first part of the excess free energy~\eqref{eq:F-azo}
is of integral form with the integrand
describing the energy costs for deviations of $\psi$ from the
equilibrium value, $\psi=0$. 
The gradient term of the energy density is taken to be 
proportional to $(\partial_z\vc{S}_{\azo})^2$, 
whereas the other term gives an increase in energy
caused by spatially uniform changes in $\psi$.
This term is written as a linear function of 
the angle dependent invariant~\eqref{eq:trace3},
$4 \Tr[\vc{S}_{\azo}^3]=3 s_{\azo}^3 \cos(3\psi)$.

For the order parameter~\eqref{eq:S-azo-surf}, 
the surface part of the energy~\eqref{eq:F-azo} can be 
represented by a quadratic polynomial of 
$\uvc{z}\cdot\vc{S}_{\azo}\cdot\uvc{z}$. 
By contrast to the elastic constant $L$
and the coefficient $B$, the surface coupling constants, $G_1$ and
$G_2$, can generally be negative leading to different boundary
conditions. For example, if $G_2=0$,
minimizing the surface term requires the $z$-component of the order
parameter, $\uvc{z}\cdot\vc{S}_{\azo}\cdot\uvc{z}=S_{zz}^{(\azo)}$,
to attain its maximal (minimal) value at the surface provided
the coefficient $G_1$ is negative (positive).
These can be referred to as the homeotropic
(planar) boundary conditions.    

The Euler-Lagrange equation for the free energy functional~\eqref{eq:F-azo} can be
readily solved to yield the relation
\begin{align}
  \label{eq:EU-sol}
  \tan(3\psi/4)=\tan(3\psi_0/4)\exp(-z/\xi),
\end{align}
where $9\xi^2=2L (B s_{\azo})^{-1}$. This relation can now be substituted
into Eq.~\eqref{eq:F-azo} to derive
the free energy as a function of the angle $\psi_0$. The result is
\begin{align}
  \label{eq:F-azo-psi0}
  F_{\azo}/h\equiv \tilde{f}_{\azo}(\psi_0)&= s_{\azo}^{5/2}\sin^2(3\psi_0/4)
+ g_1 s_{\azo}\cos(\psi_0+\pi/3)
\notag
\\
&
+g_2 s_{\azo}^2 \cos^2(\psi_0+\pi/3),
\end{align}
where $h=4(2B L)^{1/2}/3 $ and $g_i=G_i/h$.
The angle $\psi$ at the surface then can be found 
as the value of $\psi_0$ that minimizes
the function~\eqref{eq:F-azo-psi0} on the interval ranged
from $-2\pi/3$ to $2\pi/3$.

Qualitatively, dependence of $\psi_0$ on the coupling constant $g_1$
can be analyzed using elementary methods. 
For $g_2\ge 0$, we find that the angle $\psi_0$ is localized within 
different intervals depending on the value of $g_1$.
These are given by
\begin{equation}
  \label{eq:psi_vs_g1}
  \begin{cases}
-\pi/3 < \psi_{0}\le 0, & g_1\le g_c^{(1)}= -g_2 s_{\azo},\\
0 < \psi_{0}\le \pi/3, & g_{c}^{(1)}<g_1\le g_c^{(2)}=g_2
s_{\azo}+\sqrt{3}/2\, s_{\azo}^{3/2},\\
\pi/3 < \psi_{0}\le 2 \pi/3, & g_{c}^{(2)}<g_1\le g_c^{(3)}=2g_2
s_{\azo}+9/8\, s_{\azo}^{3/2},\\
\psi_{0}= 2\pi/3, & g_1> g_{c}^{(3)}.\\
\end{cases}
\end{equation}
The end points of the intervals in Eq.~\eqref{eq:psi_vs_g1}, $\psi_0=k\pi/3$ with $-1 \le k\le 2$,
represent the uniaxially anisotropic structures at the surface
\begin{align}
\label{eq:surf_struct-pi/3}
\vc{S}_{\azo}(0)=s_{\azo} (3\,\uvc{z}\otimes\uvc{z}-\vc{I})/2,
&
\quad  
\psi_0=-\pi/3,
\\
\label{eq:surf_struct-0}
\vc{S}_{\azo}(0)=-s_{\azo} (3\,\uvc{x}\otimes\uvc{x}-\vc{I})/2,
&
\quad  
\psi_0=0,
\\
  \label{eq:surf_struct+pi/3}
\vc{S}_{\azo}(0)=s_{\azo} (3\,\uvc{y}\otimes\uvc{y}-\vc{I})/2,
&
\quad  
\psi_0=\pi/3,
\\
\label{eq:surf_struct+2pi/3}
\vc{S}_{\azo}(0)=-s_{\azo} (3\,\uvc{z}\otimes\uvc{z}-\vc{I})/2,
&
\quad  
\psi_0=2\pi/3.
\end{align}
From Eqs.~\eqref{eq:surf_struct-pi/3} and~\eqref{eq:surf_struct+pi/3}
it is seen that, for the angles $\psi_0=-\pi/3$ and $\psi_0=\pi/3$,  
surface alignment will be homeotropic and homogeneous (monostable
planar), respectively. The structure~\eqref{eq:surf_struct-0} coincides with
uniaxial ordering in the bulk~\eqref{eq:dichroic2} and the surface
order parameter tensor~\eqref{eq:surf_struct+2pi/3} corresponds
to planar (random in-plane) alignment.

According to Eq.~\eqref{eq:psi_vs_g1}, the case of planar alignment
occurs only if the coupling constant $g_1$ is positive and the bulk
order parameter $s_{\azo}$ is below its critical value $s_c$
defined by the relation
\begin{align}
  \label{eq:S_c}
  g_1=2g_2 s_{c}+9/8\, s_{c}^{3/2}.
\end{align}
Fig.~\ref{fig:sc-g} shows that the critical order parameter $s_c$
is an increasing function of $g_1$. 

In Fig.~\ref{fig:sii-sa}(a) we have plotted the curves representing 
the components of the order parameter tensor at the surface
in relation to the order parameter in the bulk to
illustrate that destruction of the planar alignment takes place in a
second order transition manner. 

However, it should be stressed that
our model becomes inapplicable in the immediate vicinity of $s_c$ where
$z$-dependence of the biaxiality parameter $P_{\azo}$ 
critically slows down.  
Actually, as it can be inferred from Fig.~\ref{fig:sp-z},
the characteristic length of spatially varying biaxiality 
diverges logarithmically as $s_{\azo}$ approaches $s_c$ from above.
Under these circumstances, the assumption that
scale of the spatial variations is much shorter than the film
thickness is no more justified. 

By contrast to the boundary conditions with $g_1>0$,
there are no second order transitions provided the coupling constant $g_1$ is negative.
At sufficiently large values of $|g_1|$, the surface ordering remains nearly
homeotropic. Otherwise, the surface order parameter 
changes smoothly with $s_{\azo}$ towards the bulk order
parameter~\eqref{eq:surf_struct-0}. From Eq.~\eqref{eq:psi_vs_g1},
for $g_1=-g_2 s_{\azo}$, difference between the order parameters
vanishes. The curves presented in Fig.~\ref{fig:sii-sa}(b) illustrate this
point.

Leaving aside a detailed discussion of what happen when the coupling
constant $g_2$ is negative, we just note that in this case the
above discussed transition will generally be first order leading to
jump-like behavior of the order parameter at the surface.  

\begin{figure*}[!tbh]
\centering
\resizebox{165mm}{!}{\includegraphics*{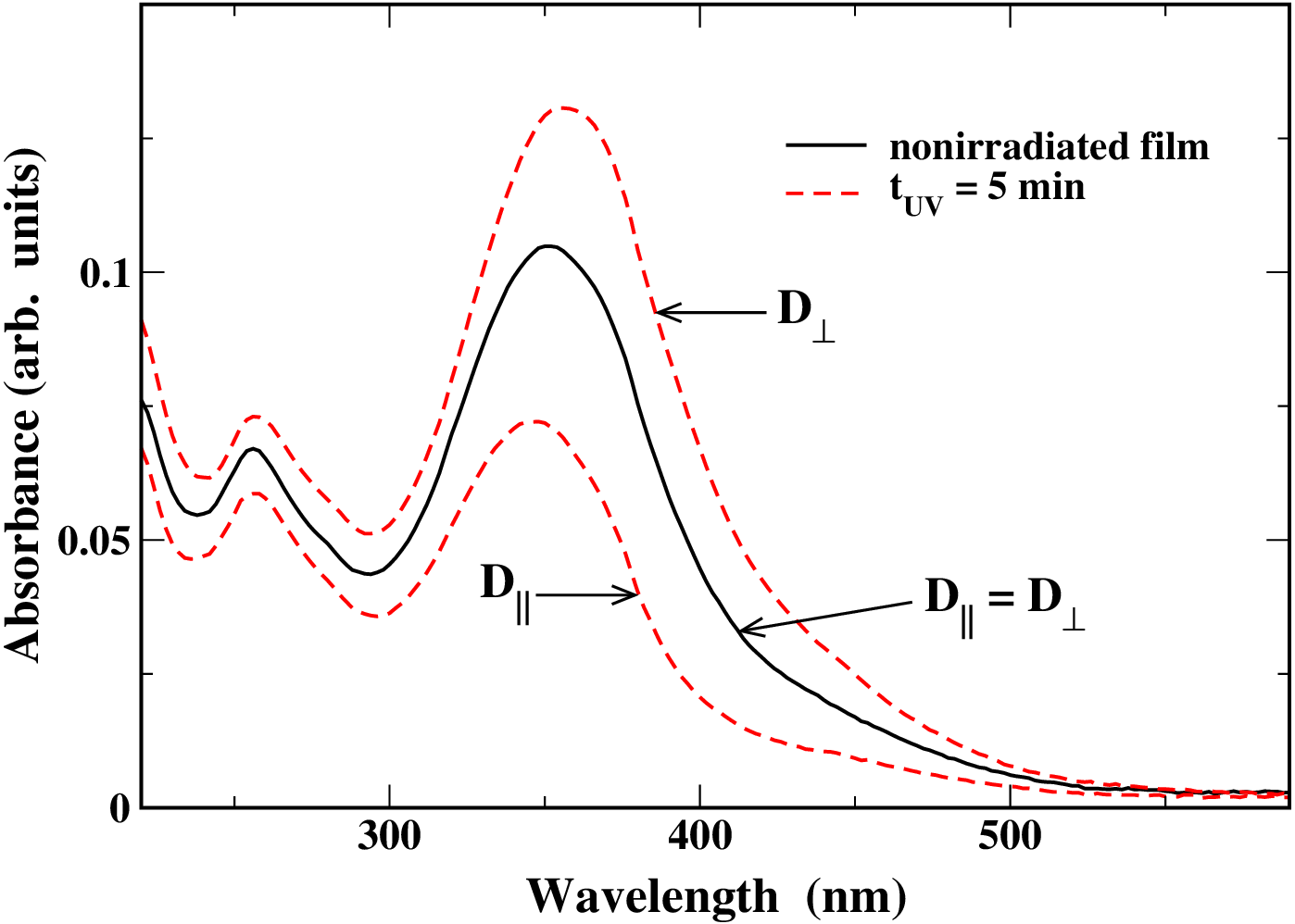}}
\caption{%
UV-visible absorption spectra of 
nonirradiated and irradiated azo-dye films.
}
\label{fig:absorb-uv}
\end{figure*}

\begin{figure*}[!tbh]
\centering
\resizebox{175mm}{!}{\includegraphics*{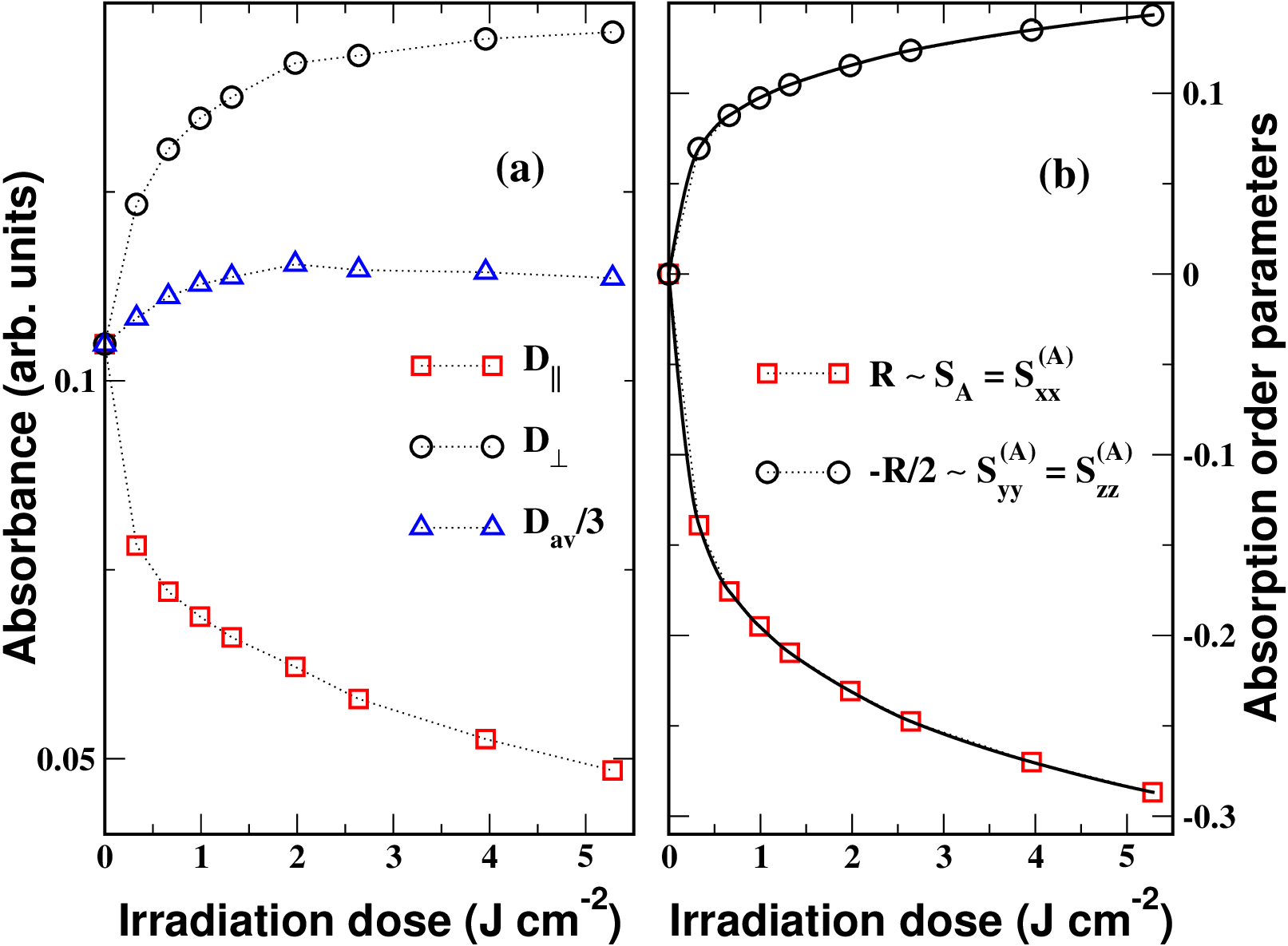}}
\caption{%
Dependence of (a)~absorption coefficients and
(b)~absorption order parameters on UV irradiation dose.
Solid lines give interpolation of the experimental data
calculated using Akima splines.
}
\label{fig:absorb-dose}
\end{figure*}

\begin{figure*}[!tbh]
\centering
\resizebox{170mm}{!}{\includegraphics*{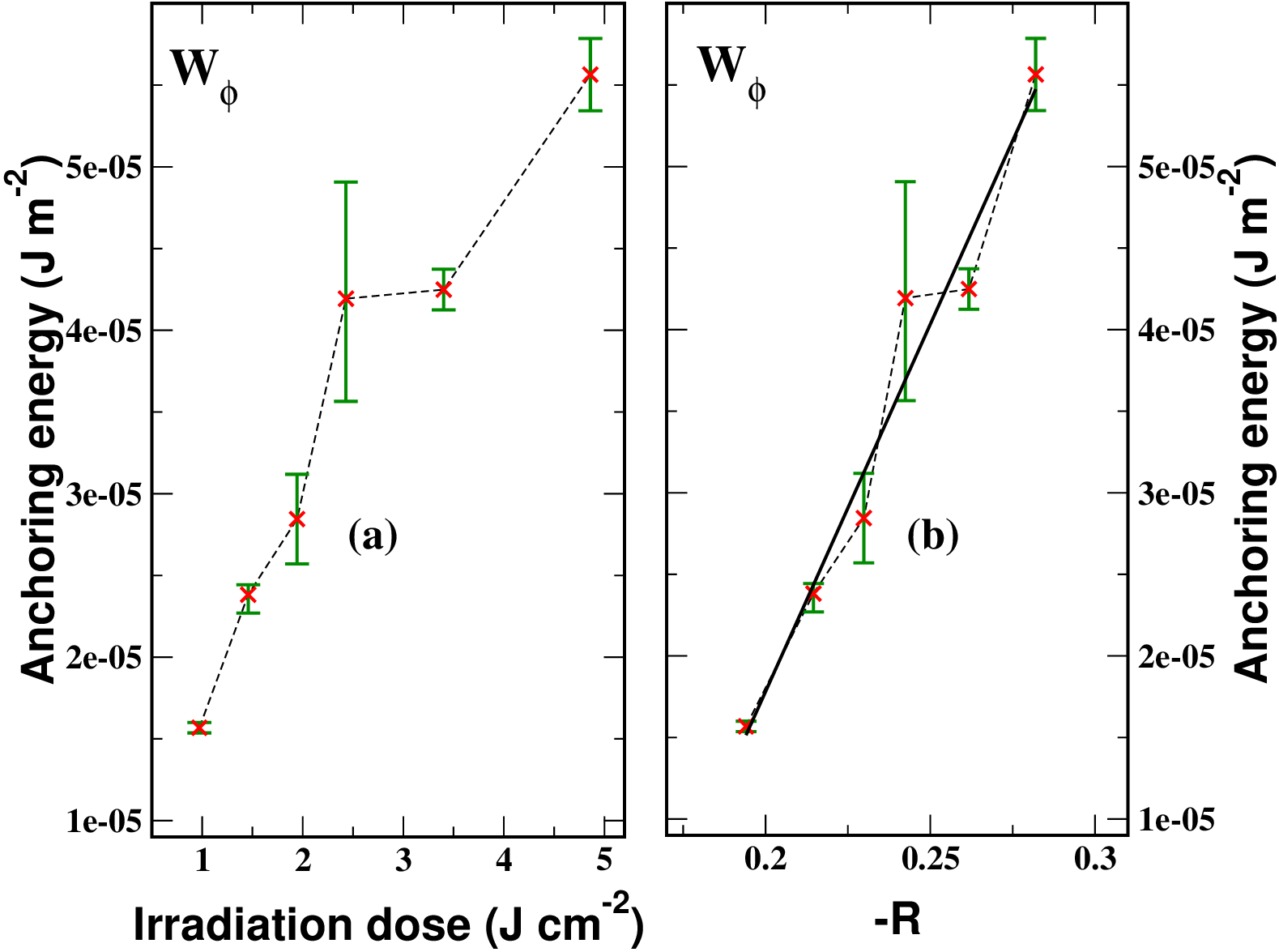}}
\caption{%
Azimuthal anchoring energy as a function of 
(a) UV irradiation dose and (b) dichroic ratio~\eqref{eq:dichroic}
which is proportional to the order parameter of azo-dye molecules,
$S_{\azo}$ (see Eq.~\eqref{eq:dichroic2}).
Solid line represents the result of linear fitting
$W_{\phi}\approx -w_a^{(0)}-w_a^{(1)} R$
with $w_a^{(0)}=0.07225$~mJ/m$^{2}$
and  $w_a^{(1)}=0.4503$~mJ/m$^{2}$.
}
\label{fig:w-a}
\end{figure*}

\begin{figure*}[!tbh]
\centering
\resizebox{175mm}{!}{\includegraphics*{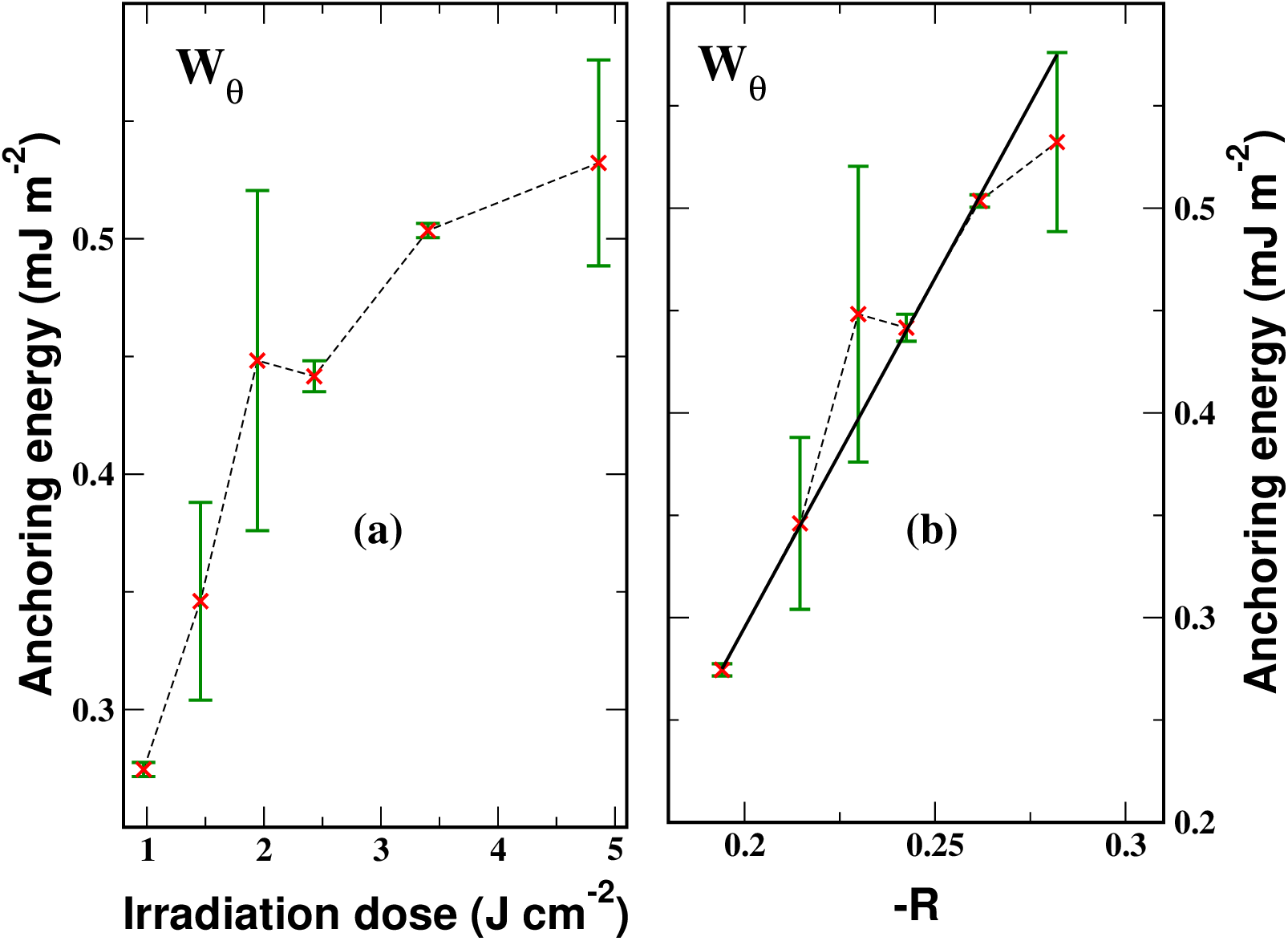}}
\caption{%
Polar anchoring energy as a function of 
(a) UV irradiation dose and (b) dichroic ratio~\eqref{eq:dichroic}
which is proportional to the order parameter of azo-dye molecules,
$S_{\azo}$ (see Eq.~\eqref{eq:dichroic2}).
Solid line represents the result of linear fitting
$W_{\theta}\approx -w_p^{(0)}-w_p^{(1)} R$
with $w_p^{(0)}=0.38714$~mJ/m$^{2}$
and  $w_p^{(1)}=3.4116$~mJ/m$^{2}$.
}
\label{fig:w-p}
\end{figure*}

\section{Experiment}
\label{sec:experim}

Now we pass on to describing the experimental procedure
employed to obtain the data linking the anchoring energy
strengths and the dichroic ratio as a measure of the photo-induced ordering.
To this end we carried out the absorption spectra and the anchoring
energy measurements for the azo-dye films irradiated at varying
exposure time. Thus, we used the samples prepared at different
irradiation doses to measure the anchoring strengths and the dichroic
ratio in relation to the dose. The data then can be recalculated to
obtain the required anchoring energy vs dichroic ratio dependence. 

\subsection{Sample preparation}
\label{subsec:sample-preparation}

Following the method
described in Ref.~\cite{Chig:lc:2002},
the azobenzene sulfuric dye SD-1 was synthesized from corresponding 
benzidinedisulfonic acid using azo coupling.
The azo-dye compound SD-1 was mixed with the polymerizable azo-dye
SDA-2 in the ratio 40\% to 60\%.
The mixture was dissolved in N,N-dimethylformamide (DMF) and
a heat initiator V-65 (from Wako pure chemical industries, Ltd.)
that was added in relation of 1:50 to SDA-2.

The solution was spin-coated onto glass substrates with
indium-tin-oxide (ITO) electrodes at 800 rpm for 5 seconds
and, subsequently, at 3000 rpm for 30 seconds.
The solvent was evaporated on a hot plate at 100\degc\  for 10 minutes.

The surface of the coated film was illuminated with linearly polarized
UV light using super-high pressure Hg lamp through an interference
filter at the wavelength 365 nm. The intensity of light irradiated on the
film surface at varying time exposure was 2.7~mW/cm$^2$.
After the photoaligning procedure, the SDA-2 films were polymerized by
heating at 150\degc\  for 1 hour in vacuum.
In order to recover quality of the photoalignment degraded after the polymerization,
the films were exposed to the UV light for 1 minute regardless of
the initial time exposure.

Two glass substrates with the photoaligned films were assembled
to form liquid crystal cells to measure the azimuthal and polar
anchoring energy strengths. The cell thickness was
5~$\mu$m and 18~$\mu$m, respectively. Liquid crystal mixtures
MLC-6080 (from Merck) in an isotropic phase 
were injected into the cell by capillary action.
 
\subsection{Absorption spectra}
\label{subsec:absorption-method}

The UV-visible absorption spectra of the films were measured in the
spectral range from 250~nm to 600~nm for the normally incident probing
light which is linearly polarized parallel (along the $x$ axis) and
perpendicular (along the $y$ axis) to the polarization vector of the
activating light.

For nonirradiated films,
the curve shown in Fig.~\ref{fig:absorb-uv} as a solid line demonstrates
that the absorption
coefficient does not depend on the polarization state of the testing
beam. By contrast, 
as it is illustrated in Fig.~\ref{fig:absorb-uv}, the absorption coefficients $D_{\parallel}$
and $D_{\perp}$ differ in the irradiated films, thus, revealing the
light induced absorption dichroism. This dichroism is mainly caused by
the photo-induced angular redistribution of the azo-dye
molecules.

By varying the exposure time the films were prepared 
at different irradiation doses and
the optical density components $D_{\parallel}$
and $D_{\perp}$ at the absorption maximum of azo-dyes
($\lambda_{m}\approx 350$~nm) were estimated from the measured absorption
spectra. The dichroic ratio then can be computed from the formula~\eqref{eq:dichroic}.
The results for the absorption coefficients and
the absorption order parameters, which are proportional
to the dichroic ratio,
are presented in Figs.~\ref{fig:absorb-dose}(a)
and~\ref{fig:absorb-dose}(b), respectively.

\subsection{Anchoring energy strengths}
\label{subsec:anchoring-energies}

The azimuthal anchoring strength, $W_{\phi}$, was measured
in a twisted nematic cell 
using the torque balance method~\cite{Iim:jap:1995,Chig:jjap:1995}.
The azo-dye aligning film and a rubbed polyimide layer
were used as confining substrates. 
The twist angle was 90 degrees. 

\begin{figure*}[!tbh]
\centering
\resizebox{165mm}{!}{\includegraphics*{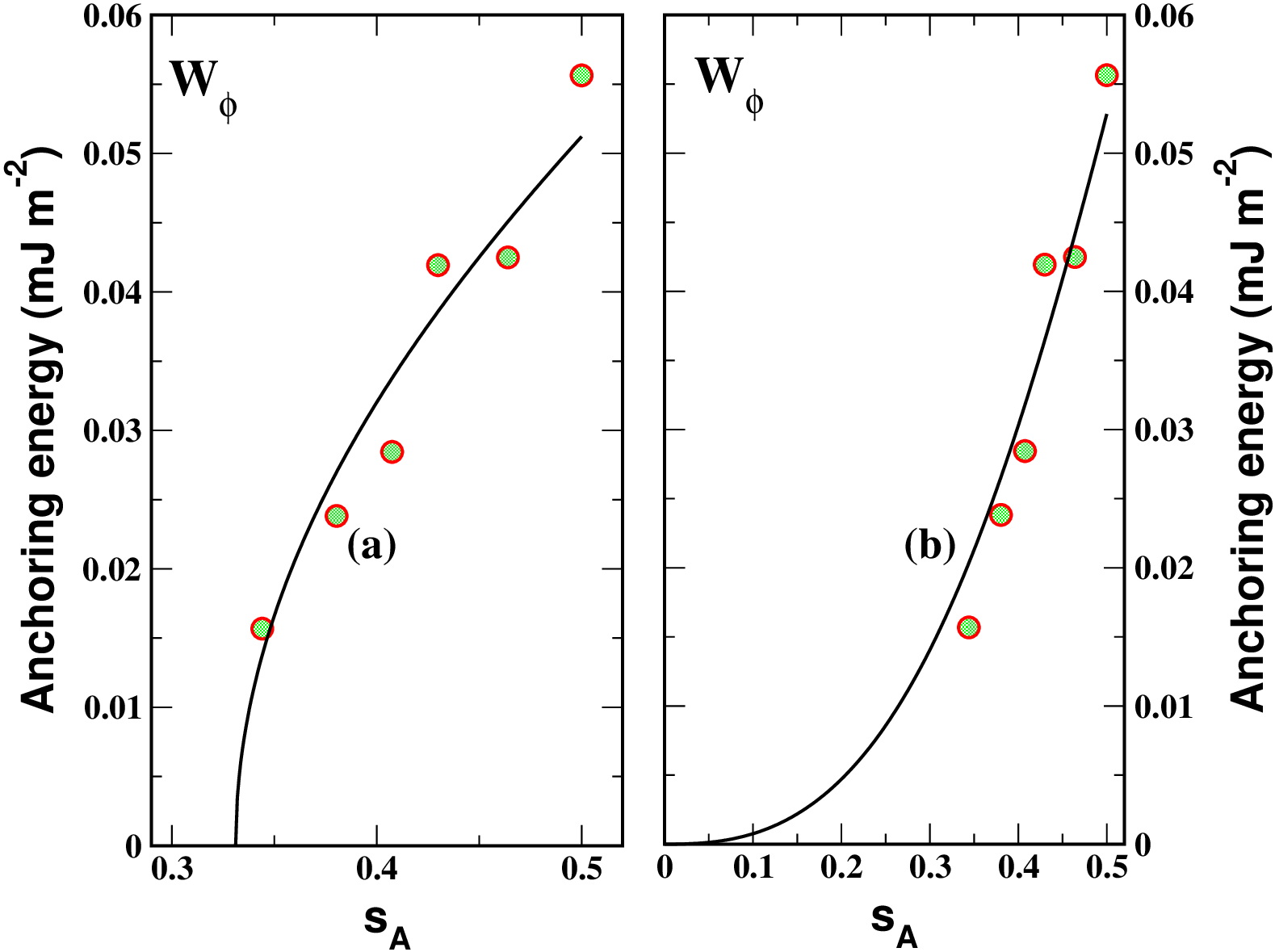}}
\caption{%
Azimuthal anchoring energy versus azo-dye order parameter.
Solid lines represent theoretical curves computed 
from the formula~\eqref{eq:wa-fit} by minimizing the
energy~\eqref{eq:F-azo-psi0} for two different sets of the parameters:
(a)~$g_1=0.47$, $g_2=0.4$, $w_{\phi}=0.067$~mJ/m$^2$;
(b)~$g_1=-0.9$, $g_2=0.4$, $w_{\phi}=0.157$~mJ/m$^2$.
}
\label{fig:wa-th}
\end{figure*}

\begin{figure*}[!tbh]
\centering
\resizebox{165mm}{!}{\includegraphics*{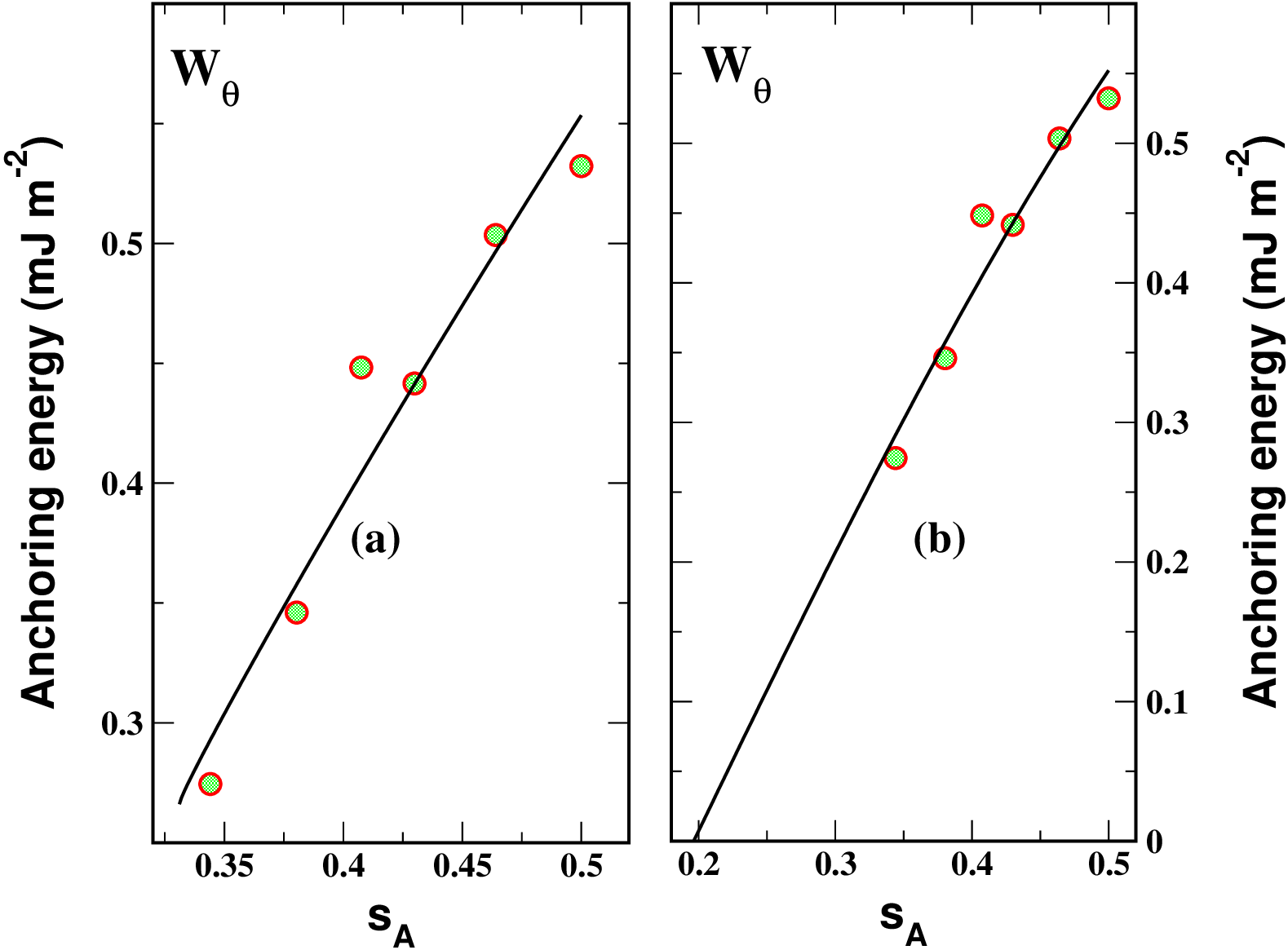}}
\caption{%
Polar anchoring energy versus azo-dye order parameter.
Solid lines represent theoretical curves computed 
from the formula~\eqref{eq:wp-fit} by minimizing the
energy~\eqref{eq:F-azo-psi0} for two different sets of the parameters:
(a)~$g_1=0.47$, $g_2=0.4$, $w_{\theta}^{(0)}=1.113$~mJ/m$^2$, 
$w_{\theta}^{(1)}=2.711$~mJ/m$^2$;
(b)~$g_1=-0.9$, $g_2=0.4$, $w_{\theta}^{(0)}=-0.382$~mJ/m$^2$, 
$w_{\theta}^{(1)}=2.247$~mJ/m$^2$.
}
\label{fig:wp-th}
\end{figure*}

Measurements of the polar anchoring strength, $W_{\theta}$,
in anti-parallel aligned cells 
were carried out using the high-voltage 
technique~\cite{Yok:jap:1985,Nasti:jap:1999,Yok:jjap:2000,Chig:pre:2003}.
The experimental data for the azimuthal and polar anchoring strengths
are plotted against the irradiation dose in Fig.~\ref{fig:wa-th}(a)
and Fig.~\ref{fig:wp-th}(a), respectively.
At low irradiation doses, when the
exposure energy is below 1~J/cm$^2$,
our experimental technique failed to provide
accurate estimates for the anchoring strengths because of
poor quality of NLC alignment in the cells.

The experimentally measured dependence of the dichroic ratio
on the irradiation dose can now be combined with the results for 
$W_{\phi}$ and $W_{\theta}$ so as
 to recalculate the anchoring energy strengths as functions 
of the dichroic ratio, $R$. The resulting data are shown in Figs.~\ref{fig:wa-th}(b)
and~\ref{fig:wp-th}(b).  

\section{Results}
\label{sec:results}

As is shown in Fig.~\ref{fig:absorb-dose}(a) representing the absorption coefficients,
$D_{\parallel}$ and $D_{\perp}$,
measured in the film irradiated  at various irradiation doses, within the
limits of experimental error, the average
absorption coefficient, $D_{\av}$, defined by the
relation~\eqref{eq:D_avr},  remains unchanged
at irradiation doses higher than 1~J/cm$^2$.
So, from the discussion given at the end of
Sec.~\ref{sec:order-param-dich} we conclude that the azo-dye
order parameter in the bulk of the film is uniaxial and is of the form
given by Eq.~\eqref{eq:dichroic2}.
In Fig.~\ref{fig:absorb-dose}(b), it is indicated that 
the components of the order parameter tensor are proportional to the
dichroic ratio~\eqref{eq:dichroic}. 
Clearly, for $D_{\perp}>D_{\parallel}$ and $\sigma_{\parallel}>\sigma_{\perp}$,
the azo-dye order parameter
$S_{\azo}$ and $R$ are both negative. 

The experimental results for
the azimuthal and polar anchoring
strengths measured in NLC cells 
with photoaligned azo-dye films used as aligning substrates
are presented in Figs.~\ref{fig:w-a}(a) and~\ref{fig:w-p}(a), respectively.
The films differ in an amount of photoinduced anisotropy which is controlled
by varying exposure time and the anchoring strengths are plotted in relation to the irradiation
dose. 

However, the fundamentally important characteristic describing 
degree of the photoinduced anisotropy is
the azo-dye order parameter. So,
in order to compare the experimental data and the theory,
we need to relate the anchoring strengths and the dichroic ratio.
Combining the anchoring energy data and
the curve depicted in Fig.~\ref{fig:absorb-dose}(b)
gives the result shown in Figs.~\ref{fig:w-a}(b) and~\ref{fig:w-p}(b).

From the relations~\eqref{eq:wa-fit}-\eqref{eq:coefp-fit}  
the anchoring strengths
depend linearly on the dichroic ratio provided the order
parameters at the surface do not differ from their values in the bulk.
The results of linear fitting of the experimental data are shown as 
solid straight lines in Figs.~\ref{fig:w-a}(b) and~\ref{fig:w-p}(b). 

Referring to Fig.~\ref{fig:w-a}(b),
the linear approximation for $W_{\phi}$ 
predicts that the azimuthal anchoring strength vanish   
at certain non-zero value of the
dichroic ratio, $R\approx -0.16$. 
By contrast, from the formulas~\eqref{eq:wa-fit}
and~\eqref{eq:coefa-fit} 
the anchoring strength $W_{\phi}$
is proportional to $R$ and, thus, disappear only in the limit of weak
photoinduced anisotropy where $R\to 0$.
Assuming that
this discrepancy can be attributed to the difference between
the bulk and surface order parameters of azo-dye,
we can apply the phenomenological model described in
Sec.~\ref{subsec:spat-vari-order} to interpret the experimental data. 
 
In the angle-amplitude representation~\eqref{eq:polar-azo},
the surface order parameters that enter the expression~\eqref{eq:wa-fit}
are given by
\begin{align}
  \label{eq:Sii-polar}
  S_{xx}^{(\azo)}\vert_{z=0}=-s_{\azo}\cos\psi_0,\quad
S_{yy}^{(\azo)}\vert_{z=0}=s_{\azo}\cos(\psi_0-\pi/3),
\end{align}
where $-s_{\azo}=S_{\azo} =(\Delta\sigma/\sigma_{\av}) R$ is the scalar order
parameter in the bulk of the azo-dye film [see
Eq.~\eqref{eq:dichroic2}].  According to our model, $\psi_0$ is the angle that minimizes 
the energy~\eqref{eq:F-azo-psi0}. 

So, the computational procedure involves two steps:
(a)~minimization of the energy~\eqref{eq:F-azo-psi0} to find the angle
$\psi_0$; and (b)~using the relations~\eqref{eq:Sii-polar} to
compute the azimuthal anchoring energy~\eqref{eq:wa-fit}.
Following this procedure, we may calculate
dependence of the anchoring strength $W_{\phi}$ on the photoinduced order
parameter $s_{\azo}$.

As it is discussed in Sec.~\ref{subsec:spat-vari-order},
the result crucially depend on the boundary conditions
that are determined by two coupling constants, $g_1$ and $g_2$. 
At $g_1\ge 0$, the surface favors planar (random in-plane) alignment
of the azo-dye molecules. In the opposite case of negative coupling
constant $g_1$, the alignment is homeotropic.

In figure~\ref{fig:wa-th}, we show the theoretical curves calculated for
both planar and homeotropic boundary conditions.
The corresponding numerical results for the polar anchoring strength
are presented in Fig.~\ref{fig:wp-th}.

The curve plotted in Fig.~\ref{fig:wa-th}(a) indicates that, for
planar boundary conditions, the azimuthal anchoring strength
$W_{\phi}$ takes non-zero values and starts growing only if the
azo-dye order parameter $s_{\azo}$ exceeds its critical value $s_c$.
Such threshold behavior is a consequence of
the second order transition discussed in Sec.~\ref{subsec:spat-vari-order}.
Contrastingly, as is shown in Fig.~\ref{fig:wa-th}(b), 
$W_{\phi}$ is a smoothly increasing function of
$s_{\azo}$ when the boundary conditions favor
the homeotropic alignment at the surface.

At first glance, the curves representing the polar anchoring strength $W_{\theta}$
plotted against $s_{\azo}$ in Figs.~\ref{fig:wp-th}(a)-(b)
do not show any noticeable differences.
It, however, should be stressed that the planar boundary conditions
prevent the polar anchoring energy from decaying to zero
as the order parameter $s_{\azo}$ decreases.
From Fig.~\ref{fig:wp-th}(b) it can be seen that
this is no longer the case when the coupling
constant $g_1$ becomes negative. 

These results show that interplay between 
photoinduced ordering in the bulk of the azo-dye films
and the preferred alignment of the azo-dye molecules at the surface
may have a profound effect on the order parameter dependence of the
anchoring energies.  
For the planar alignment with $g_1>0$, our model
predicts that the surface ordering may change
through the second order transition as the photoinduced anisotropy increases.
This transition bears close resemblance to the second order transitions in nematic
liquid crystals previously studied in Refs.~\cite{Sheng:pra:1992,Sheng:pre:1997}.

Quantitatively, the polar anchoring energy $W_{\theta}$
appears to be an order of magnitude higher than the azimuthal energy
$W_{\phi}$. From our estimates, the ratio of  
the coefficients $w_{\theta}^{(1)}$ and $w_{\phi}$,
$w_{\theta}^{(1)}/w_{\phi}=-(c_{\azo}^{(2)}+c_{\azo}^{(3)})/c_{\azo}^{(1)}$,
is likely to be well above 10.

The coefficient $c_{\azo}^{(1)}$ is negative and its sign is determined
by the dominating contribution from the Maier-Saupe term $v_{220}$
of the spherical
harmonics expansion for the azo-dye~--~NLC intermolecular potential $U_{\azo-\lc}$.
From Eq.~\eqref{eq:beta-1} the absolute value of $c_{\azo}^{(2)}$
can be significantly reduced when
the quadrupole term $v_{224}$ and the harmonics $v_{222}$
are predominately positive, so that
$\int_0^{\infty}z v_{224}(z)\, \dd z>0$ and $\int_0^{\infty}z v_{222}(z)\, \dd z>0$. 

Under these circumstances, the condition~\eqref{eq:inequal},
that requires the sum of $c_{\azo}^{(2)}$ and $c_{\azo}^{(3)}$ to be positive,
can be satisfied only if the contribution of the quadrupole term to the sum
$c_{\azo}^{(2)}+c_{\azo}^{(3)}$ is dominating. 
Thus, we may conclude that the quadrupole term
is of vital importance for the understanding 
the reasons behind the significant difference in magnitude
between the photoinduced parts of the polar and azimuthal anchoring strengths. 

\section{Discussion and Conclusions}
\label{sec:discussion}

In this paper we have studied both theoretically and experimentally
effects of photo-induced ordering in the azo-dye aligning films on the
anchoring energy strengths. These effects are governed
by dependence of the strengths on the azo-dye order parameter.

Our theoretical approach relies on the mean field theory~\cite{Sluck:jcp1:1992}
and provides general expressions for the Landau-de~Gennes surface
free energy~\eqref{eq:f_s} and the anchoring energy~\eqref{eq:f_anch}. 
The theoretical results for the azimuthal and polar anchoring energy strengths 
are obtained under certain simplifying assumptions and used to
interpret the experimental data relating the anchoring strengths and
the dichroic ratio.

We found that linear fitting of the data for the azimuthal anchoring
strength, though giving good results, predicts, in contradiction to the
bare anchoring theory, the effect of vanishing anchoring which occurs 
at certain non-zero value of the dichroic ratio (and, thus, the
irradiation dose). By using a simple phenomenological model we have
shown that this effect can be attributed to an interplay between the
light induced ordering in the bulk and the boundary conditions at the surface of
the film which may counteract the action of light. 
Thus, the bulk and surface values of the azo-dye order parameter  are
generally different.

For planar boundary conditions that favor the random in-plane alignment
of the azo-dye molecules, our model predicts threshold behavior of the azimuthal
anchoring strength  that starts growing provided the bulk order
parameter, $s_{\azo}$, exceeds its critical value. 
When the boundary conditions are homeotropic, this is no longer
the case and the azimuthal strength smoothly increases with
the dichroic ratio.
  
The results presented in Figs.~\ref{fig:wa-th} and~\ref{fig:wp-th}
demonstrate that 
the theoretical curves calculated for both types of the boundary conditions
can fit the experimental data equally well. There are, however,
differences concerning the polar anchoring strength.
By contrast to the homeotropic boundary conditions,
it never equals zero at the planar conditions.
From the previously published
results~\cite{Chig:pre:2003,Chig:pre:2005} 
it can be concluded that the homeotropic boundary conditions is unlikely to
occur in the azo-dye films under consideration.

Our final remark concerns some of the simplifying assumptions taken in
our theoretical analysis. A more sophisticated theory
that goes beyond the Fowler approximation~\eqref{eq:Fowler} is
required to take into account surface adsorption phenomena.
A self-consistent treatment of two order parameter tensors in the
interfacial layer  also remains a challenge. We hope that our results
will stimulate further progress in the field.  
 
\begin{acknowledgments}
This research was partially supported by RGC grants HKUST6102/03E
and HKUST6149/04E.
A.D.K. is indebted to Professor T.J.~Sluckin for numerous useful
conversations and acknowledges partial support from INTAS under grant
No.~03-51-5448. 
We are also grateful to Professor H.-S.~Kwok for stimulating comments.
 \end{acknowledgments}

\appendix

\section{Irreducible tensors, order parameter and invariants}
\label{sec:order-tensor}

In this appendix we introduce notations and definitions used throughout
the paper. In addition, we express the order parameter in terms of
irreducible tensors and deduce a number of algebraic relations
simplifying the derivation of the tensorial form of the intermolecular
potential given in Appendix~\ref{sec:spher-harm-expans}.

The irreducible tensors, $\vc{T}_m$, with the azimuthal number $m$ ranged
from $-2$ to 2 can be defined as linear combinations of the following form~\cite{Bie}:
\begin{equation}
  \label{eq:irr-tens}
  \vc{T}_m=\sum_{\mu, \nu = -1}^{1} C_{\mu\, \nu\, m}^{1\, 1\, 2}\, \vc{e}_\mu\otimes\vc{e}_{\nu},
\end{equation}
where $C^{\,j_1\, j_2\; j}_{m_1 m_2 m}$ is the Clebsch-Gordon
(Wigner) coefficient and
$\vc{e}_{\pm 1}=\mp(\uvc{x}\pm i\uvc{z})/\sqrt{2}$,
$\vc{e}_0=\uvc{z}$ are the vectors of spherical basis
($\uvc{x}$, $\uvc{y}$ and $\uvc{z}$ are the unit vectors directed
along the corresponding coordinate axes).
Substituting the values of the Wigner coefficient into 
Eq.~\eqref{eq:irr-tens} gives the expressions for $\vc{T}_m$
\begin{align}
  \label{eq:irr-tens-0}
&
  \vc{T}_0=(3\,\vc{e}_0\otimes\vc{e}_0-\vc{I})/\sqrt{6},
\\
\label{eq:irr-tens-1}
&
\vc{T}_{\pm 1}=(\vc{e}_0\otimes\vc{e}_{\pm 1}+
\vc{e}_{\pm 1}\otimes\vc{e}_0)/\sqrt{2},
\\
&
\label{eq:irr-tens-2}
\vc{T}_{\pm 2}=\vc{e}_{\pm 1}\otimes\vc{e}_{\pm 1},
\end{align}
so that it is not difficult to verify the validity of the
orthogonality relation
\begin{equation}
  \label{eq:orth-T}
  \Tr[\vc{T}_m \vc{T}_{-n}]=(-1)^m \delta_{mn}
\end{equation}
and the algebraic identities 
\begin{equation}
  \label{eq:alg-rel}
  \uvc{z}\cdot\vc{T}_m\cdot\uvc{z}=\sqrt{2/3}\,\delta_{m0},
\quad
  \uvc{z}\cdot\vc{T}_m\vc{T}_{-n}\cdot\uvc{z}=c_{|m|}\delta_{mn},
\end{equation}
where $c_0=2/3$, $c_1=-1/2$ and $c_2=0$.

Under the action of rotation the vectors of spherical basis transform
as follows
\begin{align}
&
  \label{eq:vect-rot}
  \vc{e}_\mu \rightarrow 
\vc{e}_\mu(\uvc{u})=\sum_{\nu=-1}^{1} D^1_{\nu \mu}(\uvc{u})\,\vc{e}_{\nu}, 
\\
&
  \label{eq:vect-0}
  \vc{e}_0(\uvc{u})=\uvc{u},
\\
&
\label{eq:vect-1}
\vc{e}_{\pm 1}(\uvc{u})=\mp (\vc{e}_x(\uvc{u})\pm i \vc{e}_y(\uvc{u}))/\sqrt{2},
\end{align}
where $D^j_{nm}(\uvc{u})\equiv D^j_{nm}(\theta,\phi)$ is the Wigner
$D$ function~\cite{Bie,Gel}; 
$\theta$ and $\phi$ are Euler angles of
the unit vector $\uvc{u}$;
$\vc{e}_x(\uvc{u})=(\cos\theta\cos\phi,\cos\theta\sin\phi,-\sin\theta)$,
$\vc{e}_y(\uvc{u})=(-\sin\phi,\cos\phi,0)$.

The definition~\eqref{eq:irr-tens} implies that 
transformation properties of the tensors $\vc{T}_m$ under rotations
are determined by the irreducible representation of the rotation group
with $j=2$, where $j$ is the angular momentum number.
So, we have 
\begin{equation}
  \label{eq:irr-tens-rot}
\vc{T}_m \rightarrow  
\vc{T}_m(\uvc{u})=\sum_{\mu, \nu = -1}^{1} C_{\mu\, \nu\, m}^{1\, 1\, 2}\, 
\vc{e}_\mu(\uvc{u})\otimes\vc{e}_{\nu}(\uvc{u})=
\sum_{k=-2}^2 D^2_{k m}(\uvc{u})\vc{T}_{k}.
\end{equation}

Eq.~\eqref{eq:irr-tens-rot} can now be combined with the
relations~\eqref{eq:vect-0} and~\eqref{eq:irr-tens-0} to yield the
expression for the order parameter tensor $\vc{Q}(\uvc{u})$:
\begin{equation}
  \label{eq:order-irr}
  \vc{Q}(\uvc{u})=\sqrt{3/2}\,\vc{T}_0(\uvc{u})
=(3\,\uvc{u}\otimes\uvc{u}-\vc{I})/2,
\end{equation}
where the unit vector $\uvc{u}$ is directed along the long molecular
axis.

The director $\uvc{n}$ is defined as an eigenvector of the
orientationally averaged order parameter tensor
\begin{equation}
  \label{eq:avr-T0}
  \avr{\vc{Q}(\uvc{u})}_{\uvc{u}}=\sqrt{3/2}\,\sum_{k=-2}^2 
\avr{D^2_{k 0}(\phi',\theta')}
\vc{T}_{k}(\uvc{n}),
\end{equation}
where $\theta'$ and $\phi'$ are Euler angles of the vector $\uvc{u}$
related to the basis vectors $\vc{e}_i(\uvc{n})$.

Since $\uvc{n}$ is the director, the averages
$\avr{D^2_{\pm 1 0}(\phi',\theta')}_{\phi', \theta'}$ vanish.
Other averages 
\begin{align}
&
  \label{eq:avr-D0}
\avr{D^2_{0 0}(\phi',\theta')}=\sqrt{2/3}\, S,
\\
&
\label{eq:avr-D2}
\avr{D^2_{\pm 2 0}(\phi',\theta')}=
P\exp(\pm 2 i \gamma)/\sqrt{6}
\end{align}
are proportional to the scalar order
parameter $S$ and the biaxiality parameter $P$.

By using the orthogonality conditions~\eqref{eq:orth-T} and
Eqs.~\eqref{eq:avr-T0}-\eqref{eq:avr-D2} we recover the relations
in the traditional form~\cite{Gennes:bk:1993}: 
\begin{align}
&
  \label{eq:avr-Q}
  \avr{\vc{Q}(\uvc{u})}_{\uvc{u}}\equiv \vc{S}(\uvc{n})=
S\,\vc{Q}(\uvc{n})+ 
P (\uvc{m}\otimes\uvc{m}-\uvc{l}\otimes\uvc{l})/2,
\\
&
\label{eq:avr-S}
S=\avr{3\,\sca{\uvc{u}}{\uvc{n}}^2-1}/2,
\\
&
\label{eq:avr-P}
P=3\avr{\sca{\uvc{u}}{\uvc{m}}^2-\sca{\uvc{u}}{\uvc{l}}^2}/2,
\end{align}
where $\uvc{m}=\cos\gamma\,\vc{e}_x(\uvc{n})-\sin\gamma\,\vc{e}_y(\uvc{n})$
and
$\uvc{l}=\sin\gamma\,\vc{e}_x(\uvc{n})+\cos\gamma\,\vc{e}_y(\uvc{n})$.

The order parameter~\eqref{eq:avr-Q} is a traceless symmetric
tensor. Therefore, there are two non-vanishing independent invariants
\begin{align}
&
  \label{eq:trace2}
  I_2=\Tr[\vc{S}^2(\uvc{n})]=\left(
3 S^2+P^2
\right)/2,
\\
&
\label{eq:trace3}
 I_3=\Tr[\vc{S}^3(\uvc{n})]=3 S\left(
S^2-P^2
\right)/4,
\end{align}
which enter the non-elastic part of the well known phenomenological
expression for the Landau-de Gennes free energy density
\begin{align}
  \label{eq:LG-density}
f_{\LG}=\frac{2 a}{3}\,(T-T^*) I_2-\frac{4 B}{3}\, I_3+\frac{4 C}{9}\,I_2^2,  
\end{align}
where $T$ is the temperature and $T^*$ is the supercooling
temperature.

For the scalar order parameters~\eqref{eq:avr-S} and~\eqref{eq:avr-P}
combined into a pair $(S, P)$, it is convenient to introduce what
might be called the ``polar'' (or amplitude-angle) representation
\begin{align}
  \label{eq:polar-rep}
  S=s\cos\psi,\quad
P=\sqrt{3}\,s\sin\psi,
\end{align}
where $s^2=2 I_2/3$.
Using the representation~\eqref{eq:polar-rep} the free energy
density~\eqref{eq:LG-density} can be recast into the form
\begin{align}
&
  \label{eq:LG-polar}
  f_{\LG}(s,\psi)=a(T-T^*) s^2- B s^3 \cos(3\psi)+ C s^4\equiv B^{2} C^{-1}
  U_{\LG},
 \\
&
\label{eq:LG-dless}
 U_{\LG}(\eta,\psi) =\frac{8+t}{32}\,\eta^2 - \eta^3\cos
 (3\psi)+\eta^4,\quad
\eta\equiv\frac{C}{B}\,s,
\end{align}
where $t=32 a C B^{-2}(T-T_c)$ is the dimensionless temperature
parameter and $T_c=T^*+B^2/(4aC)$ is the temperature of the bulk
nematic-isotropic transition.
The rescaled density~\eqref{eq:LG-dless} is a generalized version of the
dimensionless free energy density previously used in Refs.~\cite{Sheng:pra:1992,Sheng:pre:1997}.

Finally, we write down the components of the order parameter
tensor~\eqref{eq:avr-S} in the polar representation
\begin{align}
  \label{eq:Sij-polar}
  S_{ij}=s\,\left[
n_in_j\cos\psi+m_im_j\cos(\psi-2\pi/3)+l_il_j\cos(\psi+2\pi/3)
\right]
\end{align}
and notice that the stationary points of the free energy density~\eqref{eq:LG-polar} 
where the angle $\psi$ is a multiple of $\pi/3$ represent uniaxially
anisotropic states. The latter immediately recovers the well known result about
uniaxial anisotropy of NLC equilibrium states~\cite{Patash}.

\section{Tensorial form of intermolecular potential}
\label{sec:spher-harm-expans}

We begin with the intermolecular potential between two rigid, axially
symmetric molecules expanded in a series of spherical harmonics
as follows~\cite{Pople:1954}
\begin{align}
&
  \label{eq:poten-gen}
  U(\vc{r},\uvc{u}_1,\uvc{u}_2)=\sum_{j_1, j_2, j} u_{j_1 j_2 j}(r)
\sum_{m_1, m_2, m} C^{\,j_1\, j_2\; j}_{m_1 m_2 m}
\notag
\\
&
\times
Y_{j_1 m_1}(\uvc{u}_1)Y_{j_2 m_2}(\uvc{u}_2)Y^{*}_{j m}(\uvc{r})
\end{align}
where $\vc{r}\equiv\vc{r}_{12}=\vc{r}_1-\vc{r}_2$, 
$\vc{r}_i$ and $\uvc{u}_i$ are the position and orientation
(equivalently, Euler angles of the long molecular axis) 
coordinates of the interacting molecules, respectively;
$Y_{jm}(\uvc{u})=\sqrt{(2j+1)/(4\pi)}D^{j\,*}_{m0}(\uvc{u})$
is the spherical function~\cite{Bie,Abr}.
The form of the expansion~\eqref{eq:poten-gen} implies that the
potential is invariant under translations, $\vc{r}_i\to
\vc{r}_i+\Delta\vc{r}$, and rotations, 
$\{\vc{r}_i, \uvc{u}_i\}\to \{R\,\vc{r}_i, R\,\uvc{u}_i\}$. 
In addition, we shall assume the
head-tail symmetry
\begin{equation}
  \label{eq:head-tail}
 U(\vc{r},\uvc{u}_1,\uvc{u}_2)=U(\vc{r},-\uvc{u}_1,\uvc{u}_2)=U(\vc{r},\uvc{u}_1,-\uvc{u}_2), 
\end{equation}
so that the outer sum in Eq.~\eqref{eq:poten-gen}
is restricted to run over even values of $j_1$ and $j_2$.

It is now our task to link the pairwise potential integrated over
in-plane coordinates
\begin{align}
&
  \label{eq:avr-poten}
  V(z,\uvc{u}_1,\uvc{u}_2)=\int_S U(\vc{r},\uvc{u}_1,\uvc{u}_2)\dd x
  \dd y
\notag
\\
&
=\sum_{j_1, j_2, j} 
[(2j_1+1)(2j_2+1)(2j+1)/(4\pi)^3]^{1/2}
\notag
\\
&
\times
v_{j_1 j_2 j}(z)\,
\sum_{m} C^{\;j_1\; j_2\; j}_{m\, -m\, 0}\,
D^{j_1}_{m 0}(\uvc{u}_1)\,D^{j_2}_{-m 0}(\uvc{u}_2)\,D^{j}_{0 0}(\uvc{z})
\end{align}
and the tensorial representation that was
originally suggested by Ronis and Rosenblatt in Ref.~\cite{Ron:pra:1980}
\begin{align}
&
  \label{eq:poten-in-Q}
  V(z,\uvc{u}_1,\uvc{u}_2)\equiv V(z,\vc{Q}_1,\vc{Q}_2)
=V_{\iso}(z)+\beta^{(0)}(z)\,\uvc{z}\cdot(\vc{Q}_1+\vc{Q}_2)\cdot\uvc{z}
\notag
\\
&
+\beta^{(1)}(z)\, \Tr(\vc{Q}_1\vc{Q}_2)
+\beta^{(2)}(z)\, \uvc{z}\cdot\vc{Q}_1\vc{Q}_2\cdot\uvc{z}
\notag
\\
&
+\beta^{(3)}(z)\, [\uvc{z}\cdot\vc{Q}_1\cdot\uvc{z}]
[\uvc{z}\cdot\vc{Q}_2\cdot\uvc{z}],
\end{align}
where $\vc{Q}_i\equiv\vc{Q}(\uvc{u}_i)$ is defined by
Eq.~\eqref{eq:order-irr}.
In other words,
the problem is to express the coefficients $\beta^{(i)}(z)$ in terms
of the harmonics $v_{j_1 j_2 j}(z)$.
To this end we restrict ourselves to the lowest order harmonics of the
expansion~\eqref{eq:avr-poten} with $j_i< 4$ and consider the equation
\begin{align}
&
  \label{eq:system-aux}
  x_1\, [\uvc{z}\cdot\vc{T}_0(\uvc{u}_1)\cdot\uvc{z}]
[\uvc{z}\cdot\vc{T}_0(\uvc{u}_2)\cdot\uvc{z}]
+x_2\, \Tr(\vc{T}_0(\uvc{u}_1)\vc{T}_0(\uvc{u}_2))
\notag
\\
&
+x_3\, \uvc{z}\cdot\vc{T}_0(\uvc{u}_1)\vc{T}_0(\uvc{u}_2)\cdot\uvc{z}
=\sum_{m=0}^2 \alpha_m D^2_{m0}(\uvc{u}_1) D^2_{-m0}(\uvc{u}_2),
\end{align}
that need to be solved for $x_1$, $x_2$ and $x_3$.
The sum on the right hand side of Eq.~\eqref{eq:system-aux}
represents sum of the harmonics with $j_1=j_2=2$.
The case of the harmonics with $j_1 j_2 = 0$ is much easier to treat
as we only have to use the relation
\begin{equation}
  \label{eq:rel-j2j1=0}
  \uvc{z}\cdot\vc{T}_0(\uvc{u})\cdot\uvc{z}=D^2_{00}(\uvc{u}).
\end{equation}
 
By using Eq.~\eqref{eq:irr-tens-rot} combined with the
relations~\eqref{eq:orth-T} and~\eqref{eq:alg-rel} it is
straightforward to transform Eq.~\eqref{eq:system-aux} into a system
of linear equations. The solution of the system is given by
\begin{align}
&
  \label{eq:sol-syst}
  x_1=(3\alpha_0 + 4\alpha_1+\alpha_2)/2,
\notag
\\
&
x_2=-2 (\alpha_1+\alpha_2),
\notag
\\
&
x_3=\alpha_2.
\end{align}

Given the values of the Wigner coefficients, we can now use the
relations~\eqref{eq:rel-j2j1=0} and~\eqref{eq:sol-syst} to derive the
final result in the following form:
\begin{equation}
  \label{eq:V-iso}
  V_{\iso}(z)=v_{000}(z)/(4\pi)^{3/2},
\end{equation}
\begin{equation}
  \label{eq:alph}
  [(4\pi)^{3/2}/5]\, \beta^{(0)}(z)= v_{202}(z)
= v_{022}(z),
\end{equation}
\begin{align}
&
  \label{eq:beta-1}
  [(4\pi)^{3/2}/5]\, \beta^{(1)}(z)= 2/(3\sqrt{5})
\bigl\{ v_{220}(z)
\notag
\\
&
+10/\sqrt{14}\, v_{222}(z)+3/\sqrt{14}\,v_{224}(z) \bigr\},
\end{align}
\begin{equation}
  \label{eq:beta-2}
  [(4\pi)^{3/2}/5]\, \beta^{(2)}(z)= -20/\sqrt{70}
\bigl\{ v_{222}(z)+v_{224}(z) \bigr\},
\end{equation}
\begin{equation}
  \label{eq:beta-3}
  [(4\pi)^{3/2}/5]\, \beta^{(3)}(z)= 35/\sqrt{70}\, v_{224}(z).
\end{equation}

The formulas~\eqref{eq:V-iso}--\eqref{eq:beta-3} relate the parameters
of the representation~\eqref{eq:poten-in-Q} to the coefficient
functions in the spherical harmonics expansion~\eqref{eq:avr-poten}.
The terms $v_{202}$ and $v_{022}$ 
describe the coupling between orientation of the molecules and the
intermolecular vector, whereas  $v_{220}$ and $v_{224}$ are known as
the Maier-Saupe and the quadrupole terms, respectively.
For the interaction between NLC molecules, the functions
$\beta^{(1)}(z)$ and $\beta^{(2)}(z)$ define the elastic coefficients
of NLC that must be positive. This stability condition implies that
$\beta^{(1)}$ and $\beta^{(2)}$ are both predominately
non-positive~\cite{Sen:1987,Sulliv:jcp:1988,Sluck:jcp1:1992,Hess:jcp:1993}.

\bibliographystyle{apsrev}
\bibliography{polymer,scatter,lc,quant,hk}

\end{document}